\documentclass[useAMS,usenatbib,fleqn]{mn2e}
\usepackage{aas_macros}
\usepackage{graphicx}
\graphicspath{{/murphy/seb_stuff/SFXT/COMB_OBS_ANALYSIS/IGRJ16418/PAPER_PLOTS/}}
\usepackage{fixltx2e}
\usepackage{amsmath}
\title[IGR J16418$-$4532]{\emph{INTEGRAL} and \emph{XMM-Newton} observations of IGR J16418$-$4532: evidence of accretion regime transitions in a supergiant fast X-ray transient}
\author[S. P. Drave et al.]
	{S. P. Drave$^1\thanks{sd805@soton.ac.uk}$, A. J. Bird$^1$, L. Sidoli$^2$, V. Sguera$^{3}$, V. A. McBride$^{4,5}$, A. B. Hill$^{7,1}$,
        \newauthor
         A. Bazzano$^6$, and M. E. Goossens$^1$  \\
        $^1$School of Physics and Astronomy, Faculty of Physical Sciences and Engineering, University of Southampton, University Road, \\ 
        Southampton, SO17 1BJ, UK \\
        $^2$INAF-IASF, Istituto di Astrofisica Spaziale e Fisica Cosmica, Via E. Bassini 15, I-20133 Milano, Italy \\
        $^3$INAF-IASF, Istituto di Astrofisica Spaziale e Fisica Cosmica, Via Gobetti 101, Bologna, Italy \\
        $^4$Department of Astronomy, Astrophysics, Cosmology and Gravity Centre, University of Cape Town, Private Bag X3 Rondebosch, 7701, \\ 
        South Africa \\
        $^5$South African Astronomical Observatory, PO Box 9, Observatory, 7935 South Africa \\
        $^6$IAPS-INAF, Istituto di Astrofisica Spaziale e Fisica Cosmica, Via del Fosso del Cavaliere 100, 00133 Roma, Italy \\
        $^7$W. W. Hansen Experimental Physics Laboratory, Kavil Institute for Particle Astrophysics and Cosmology, Department of Physics, and  \\
        SLAC National Accelerator Laboratory, Stanford University, Stanford, CA 94305, USA \\
        }

\date{Accepted 2013 April 27.  Received 2013 April 26; in original form 2013 Febuary 19}

\pagerange{\pageref{firstpage}--\pageref{lastpage}} \pubyear{2013}

\def\LaTeX{L\kern-.36em\raise.3ex\hbox{a}\kern-.15em
    T\kern-.1667em\lower.7ex\hbox{E}\kern-.125emX}

\begin{document}

\label{firstpage}

\maketitle

\begin{abstract}
We report on combined \emph{INTEGRAL} and \emph{XMM-Newton} observations of the supergiant fast X-ray transient IGR J16418$-$4532. The observations targeted the X-ray eclipse region of IGR J16418$-$4532s orbit with continuous \emph{INTEGRAL} observations across $\sim$25\% of orbital phase and two quasi-simultaneous \emph{XMM-Newton} observations of length 20\,ks and 14\,ks, occurring during, and just after the eclipse respectively. An enhanced \emph{INTEGRAL} emission history is provided with 19 previously unreported outbursts identified in the archival 18$-$60\,keV data set. The \emph{XMM-Newton} eclipse observation showed prominent Fe-emission and a flux of 2.8$\times$10$^{-13}$\,erg\,cm$^{-2}$\,s$^{-1}$ (0.5 - 10 keV). Through the comparison of the detected eclipse and post eclipse flux, the supergiant mass loss rate through the stellar wind was determined as $\dot{M}_{w}$\,$=$\,2.3-3.8$\times$10$^{-7}$\,M$_{\odot}$\,yr$^{-1}$. The post eclipse \emph{XMM-Newton} observation showed a dynamic flux evolution with signatures of the X-ray pulsation, a period of flaring activity, structured n$_{H}$ variations and the first ever detection of an X-ray intensity dip, or `off-state', in a pulsating supergiant fast X-ray transient. Consideration is given to the origin of the X-ray dip and we conclude that the most applicable of the current theories of X-ray dip generation is that of a transition between Compton cooling dominated and radiative cooling dominated subsonic accretion regimes within the `quasi-spherical' model of wind accretion. Under this interpretation, which requires additional confirmation, the neutron star in IGR J16418$-$4532 possesses a magnetic field of $\sim$10$^{14}$\,G, providing tentative observational evidence of a highly magnetised neutron star in a supergiant fast X-ray transient for the first time. The implications of these results on the nature of IGR J16418$-$4532 itself and the wider SFXT class are discussed.  
\end{abstract}


\section{Introduction}

Supergiant Fast X-ray Transients (SFXT) are a sub-class of supergiant high mass X-ray binaries (HMXB) that display extreme flaring behaviour on short ($\sim$ hour) timescales \citep{2005A&A...444..221S}. They also display an X-ray dynamic range in excess of that possessed by classical supergiant X-ray binaries ($\sim$\,10$-$20, Sg-XRB), reaching 10$^{4}$ $-$10$^5$ in the most extreme systems such as IGR J17544$-$2619 \citep{2009ApJ...707..243R} whilst only reaching about 10$^2$ in the so-called intermediate SFXTs, such as IGR J16465$-$4514 \citep{2010MNRAS.406L..75C}. Due to the supergiant nature of their companion stars \citep{2006ESASP.604..165N} SFXTs are located along the Galactic Plane where there are currently 10 spectroscopically confirmed systems \citep{2011arXiv1111.5747S}, as well as a similar number of candidate SFXTs that show the required X-ray flaring behaviour but, as yet, do not have a spectroscopically confirmed supergiant counterpart. 

IGR J16418$-$4532 was first detected by \emph{INTEGRAL} during observations of the transient black hole system 4U 1630$-$47 \citep{2004ATel..224....1T} and proposed as a member of the SFXT class by \citet{2006ApJ...646..452S} after the identification of short duration outbursts from the system. \emph{XMM-Newton} observations reported by \citet{2006A&A...453..133W} identified a neutron star pulse period of 1246\,$\pm$\,100\,s (later refined to 1212\,$\pm$6\,s by \citealt{2012MNRAS.420..554S}) along with a high ($\sim$ 10$^{23}$ cm$^{-2}$) absorption intrinsic to the system. The source was also localised to arcsecond accuracy and the IR counterpart proposed as 2MASS J16415078$-$4532253, which was later confirmed by \citet{2012MNRAS.419.2695R} through a \emph{Swift}/XRT source localisation of RA,Dec (J2000) $=$ 16$^{h}$ 41$^{m}$ 50$^{s}$.65, -45$^{o}$ 32$^{\prime}$ 27$^{\prime\prime}$.3 with an uncertainty of 1$^{\prime\prime}$.9 (90\% confidence). \citet{2006ATel..779....1C} identified a short, eclipsing orbit of 3.7389$\pm$0.0004\,days and 3.753$\pm$0.004\,days in the system using \emph{RXTE}/ASM and \emph{Swift}/BAT data sets respectively\footnote{the statistical discrepancy in these two values is likely due to an underestimate of either one or both uncertainties} (this determination was recently improved to 3.73886 $\pm$ 0.00028\,days using \emph{RXTE}/ASM data over an $\sim$ 14\,year baseline \citep{2011ApJS..196....6L}). Combining the orbital and pulse periods places IGR J16418$-$4532 in the wind-fed Sg-XRB region of the Corbet diagram \citep{1986MNRAS.220.1047C}. \citet{2008A&A...484..783C} performed NIR photometry on the IR counterpart of IGR J16418$-$4532 and, using SED fitting, determined a stellar temperature of 32,800\,K suggesting the companion is of spectral type OB. Along with the location of IGR J16418$-$4532 in the Corbet diagram, the SED results confirmed the HMXB nature of this system and implied a minimum distance of $\sim$ 13\,kpc when assuming a supergiant companion. The supergiant nature of the companion was recently confirmed by Coleiro et al. (2013, private communication) who classified the companion as a BN0.5Ia nitrogen rich supergiant, thus confirming IGR J16418$-$4532 as a intermediate SFXT.  

The maximal X-ray dynamic range observed in IGR J16418$-$4532, greater than 10$^{2}$ in the soft X-ray regime (\citealt{2012MNRAS.419.2695R}, \citealt{2012MNRAS.420..554S}), is in excess of that observed from classical, wind-fed Sg-XRBs with comparable orbital parameters. Through modelling the flare - luminosity distribution observed during an intensive \emph{Swift}/XRT monitoring campaign of IGR J16418$-$4532, \citet{2012MNRAS.419.2695R} concluded that the enhanced level of variability in this system could be described by the NS accreting stellar wind clumps in the mass range 10$^{16}$ - 10$^{21}$\,g embedded in a highly structured wind with a terminal velocity between 800 and 1300\,km s$^{-1}$ (using the \citet{2009MNRAS.398.2152D} formulation of the `clumpy wind' scenario of SFXT outbursts (\citealt{2005A&A...441L...1I}, \citealt{2007A&A...476..335W})). However using a 40\,ks \emph{XMM-Newton} observation in 2011, \citet{2012MNRAS.420..554S} argue that the observed X-ray variability, orbital dynamics and quasi-periodic flaring activity is indicative of the NS accreting in a regime that is transitional between pure wind accretion and full Roche lobe overflow (RLO). It is argued that such a `Transitional Roche Lobe Overflow (TRLO)' regime could be the dominant source of the observed X-ray variability in SFXTs with short orbital periods. 

In this work we present new quasi-simultaneous, orbital phase targeted \emph{INTEGRAL} and \emph{XMM-Newton} observations, along with an archival \emph{INTEGRAL} study, of IGR J16418$-$4532. In Section \ref{sect:IDR} we outline the data sets, analysis and results from both the archival and new \emph{INTEGRAL} observations. Section \ref{sect:XDR} presents the analysis of, and results from, the new \emph{XMM-Newton} data set. These results are then discussed in Section \ref{sect:DISC} and final conclusions drawn in Section \ref{sect:CONC}.

\section{\emph{INTEGRAL} data analysis and results}

The archival data set from \emph{INTEGRAL}/IBIS (\citealt{2003A&A...411L...1W}, \citealt{2003A&A...411L.131U}) consisted of all observations of IGR J16418$-$4532 spanning 2003 January 11 through 2010 September 30 providing a total exposure of $\sim$10\,Ms. All observations were processed with version 9 of the \emph{INTEGRAL} Offline Science Analysis software (OSA, \citealt{2003A&A...411L.223G}) and images were created in the 18$-$60\,keV energy range for each science window (ScW, individual \emph{INTEGRAL} exposures of nominal length 2000\,s). An 18$-$60\,keV light curve was generated by extracting count rates and errors from each image at the best determined X-ray position of IGR J16418$-$4532. Our new \emph{INTEGRAL} observations were performed between UTC 09:57:33 2012-09-01 and 09:00:29 2012-09-02 for a total exposure of $\sim$78\,ks. Images were again generated in the 18$-$60\,keV band and a light curve extracted following the same procedures outlined for the archival data.

The light curve was tested for periodicities by constructing a Lomb-Scargle periodogram (\citealt{1976Ap&SS..39..447L}, \citealt{1982ApJ...263..835S}) and the peak period was identified as 3.74\,days. The uncertainty on the peak was derived via a Monte-Carlo based randomisation test whereby the count rate of each point in the light curve is randomly varied within its error bar using a Gaussian probability distribution. A periodogram is then generated for the altered light curve and the strongest period recorded. This process is repeated 50,000 times and the resulting distribution of peak periods fit with a Gaussian curve, the width of which is taken as the 1$\sigma$ error on the identified periodicity. The periodicity identified in the archival data set is therefore 3.7399 $\pm$ 0.0003\,days. This value is consistent, within 2$\sigma$, with the orbital period of 3.73886 $\pm$ 0.00028\,days reported by \citet{2011ApJS..196....6L}, which was identified in an \emph{RXTE}/ASM data set spanning 14 years. As the \citet{2011ApJS..196....6L} determination has a slightly better constraint and originates from a longer data set it is this value of the orbital period that will be used for the remainder of this work. Additionally to maintain consistency with previous works the zero phase ephemeris of MJD 53560.200 \citep{2006ATel..779....1C} will be used throughout.

\begin{figure}
%\begin{center}
	\includegraphics[width=0.5\textwidth]{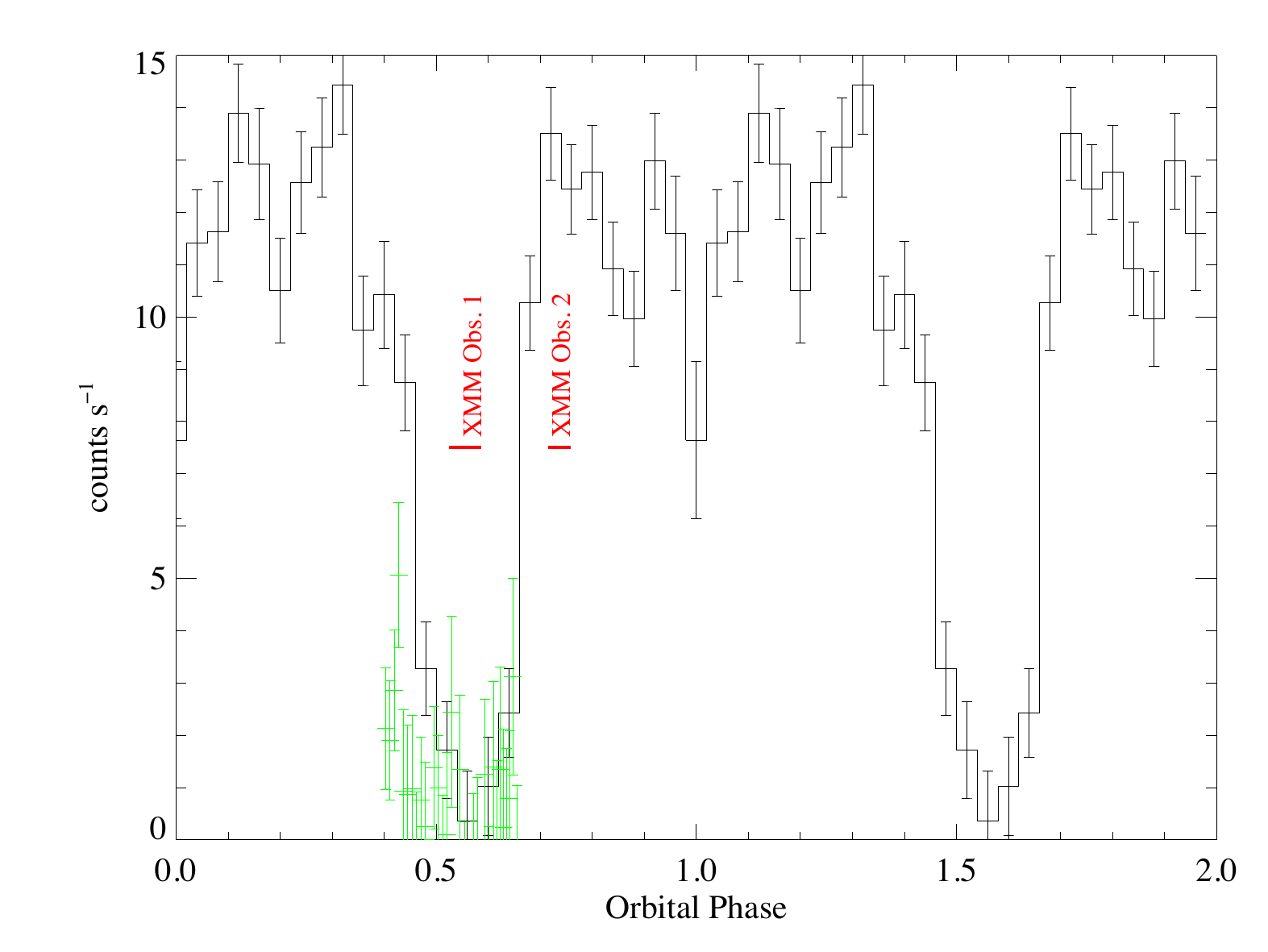}
	\caption{Archival \emph{INTEGRAL}/IBIS 18 $-$ 60\,keV light curve (scaled by a factor of 10) folded on P$_{orb}$ $=$ 3.73886\,days with the zero phase ephemeris MJD 53560.20. The new IBIS observations are shown by the green points and the phase locations of the \emph{XMM-Newton} observations are shown in red.}
\label{fig:pfold}
%\end{center}
\end{figure}

Figure \ref{fig:pfold} shows the archival 18 $-$ 60\,keV IBIS light curve folded on the 3.73886\,day orbital period and scaled by a factor of 10 for illustrative purposes. Additionally the 18 - 60 keV ScW light curve of the new \emph{INTEGRAL} observations is over plotted in green and the orbital phase locations of the \emph{XMM-Newton} observations are shown in red (see below for further details). The orbital profile is dominated by the deep eclipse that spans $\sim$0.2 of the phase space between $\phi$ $=$ 0.45 and 0.65, corresponding to a duration of $\sim$0.75\,days, and is consistent with the neutron star being fully eclipsed. Outside of the eclipse, the orbital profile is consistent with a constant flux at an unscaled IBIS count rate of $\sim$1.2\,counts per second, corresponding to a flux of 7 mCrab in the 18 $-$ 60 keV band, showing that the eclipse is the driving factor in the observed periodicity. 

The archival 18 $-$ 60\,keV light curve was also searched for outbursts by systematically identifying regions of the light curve with a high local significance in windows of 0.02 to 3.0\,days ($\sim$half an hour to the majority of one orbit). Due to the coded aperture nature of IBIS and the deconvolution method of image reconstruction, light curves of transient sources contain a high amount of systematic noise that is symmetric about zero during times of non-detection. To guard against false outburst detections resulting from the large number of trials incurred by this method of outburst identification, the count rates in the light curve are inverted about zero and the same identification procedure performed. The maximum significance detected in the inverted light curve, which is a false positive by definition, is then taken as the cut-off point below which the local significances can not be safely distinguished from random statistical variations present in the light curve. Applying this cut-off to the original set of excesses produced 35 distinct outburst events within the archival light curve down to a minimal significance of 4.48$\sigma$ and with durations in the range $\sim$0.02 to 1.3\,days. 16 of the 35 detected outbursts have already been reported in previous works (see \citealt{2006ApJ...646..452S}, \citealt{2010MNRAS.408.1540D}) while the remaining 19 are new detections, the main properties of which are reported in Table \ref{tab:IBISbursts}. 

\begin{figure}
	\includegraphics[width=0.5\textwidth]{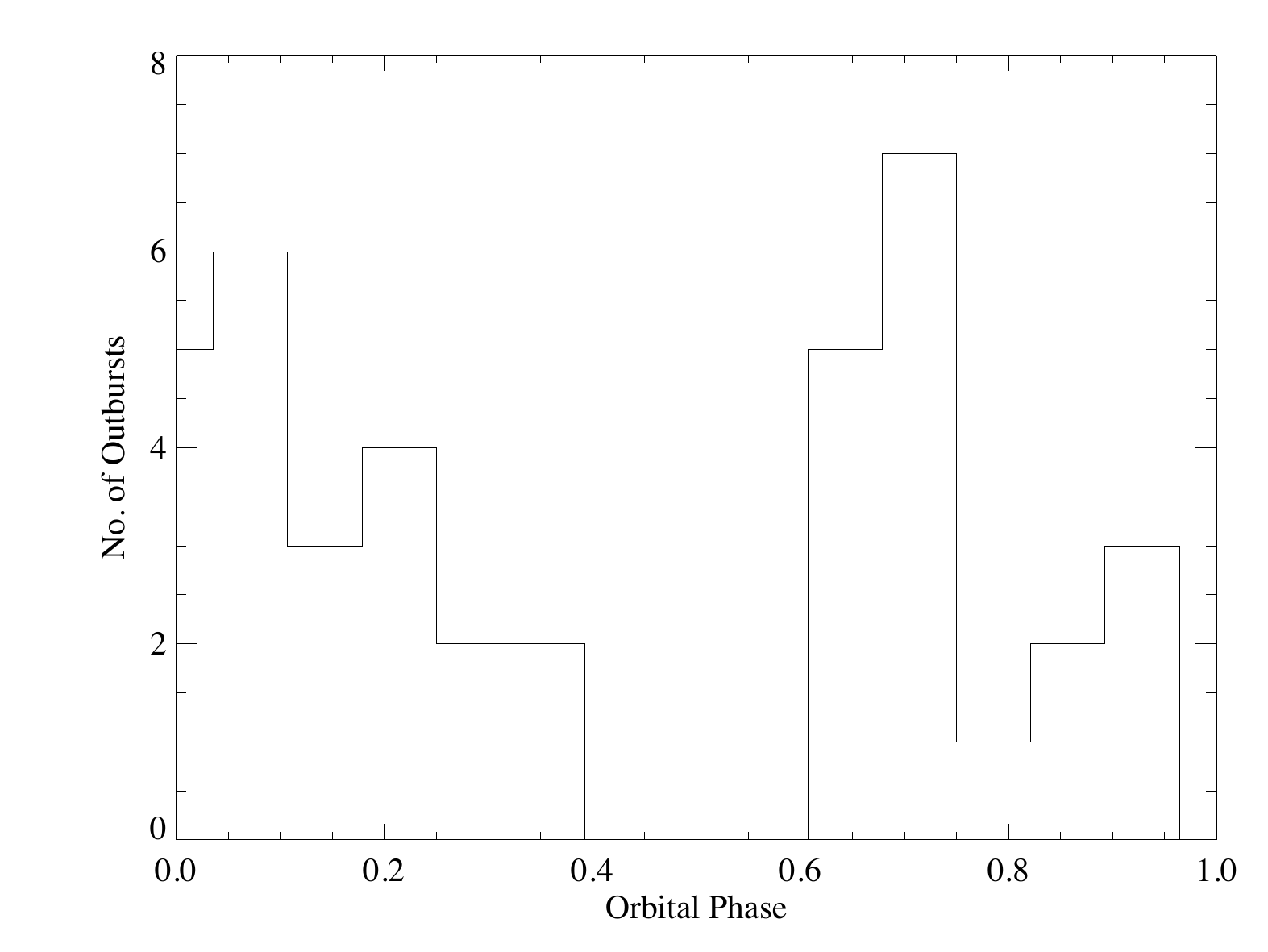}
	\caption{Orbital Phase Distribution of the archival IGR J16418$-$4532 outbursts using the ephemeris P$_{orb}$ $=$ 3.73886\,days and a zero phase of MJD 53560.20.}
\label{fig:burstphase}
\end{figure}

The total duration of these outbursts places the active duty cycle of IGR J16418$-$4532 at 6.14\%. This figure is somewhat higher than that derived from the results of \citet{2010MNRAS.408.1540D} which give a duty cycle of $\sim$1.3\%. Whilst the analyses were performed on data from different energy bands (18\,$-$\,60\,keV here and 20\,$-$\,40\,keV in \citealt{2010MNRAS.408.1540D}) and on data sets of different total exposure, it is likely the difference in the methods used to identify outbursts in each work is the main cause of this disparity. \citet{2010MNRAS.408.1540D} consider only the ScWs in which IGR J16418$-$4532 was detected at a significance of $>$\,5$\sigma$ from which to define outbursts whereas here we consider all ScWs in our searches and model the outburst significance cutoff to apply as described above. Under this method an outburst search becomes sensitive to a population of longer, fainter outbursts which are below the formal detection threshold for a known source in a single ScW (i.e. 5$\sigma$) but sum to a significant detection when the multiple ScWs of an outburst are combined. The extent to which the size of the increase in the active duty cycle observed for IGR J16418$-$4532 is specific to this source or represents a more global phenomenon however requires further investigation. Figure \ref{fig:burstphase} shows the distribution, in orbital phase, of the peak emission from each identified outburst. It can be seen that, apart from the eclipse region where no outbursts are detected, there is no coherent structure in the outburst distribution which is consistent with the phase folded light curve (Fig. \ref{fig:pfold}). This implies that outbursts are equally likely to occur at any point in the orbital phase.

\begin{figure}
	\includegraphics[width=0.5\textwidth]{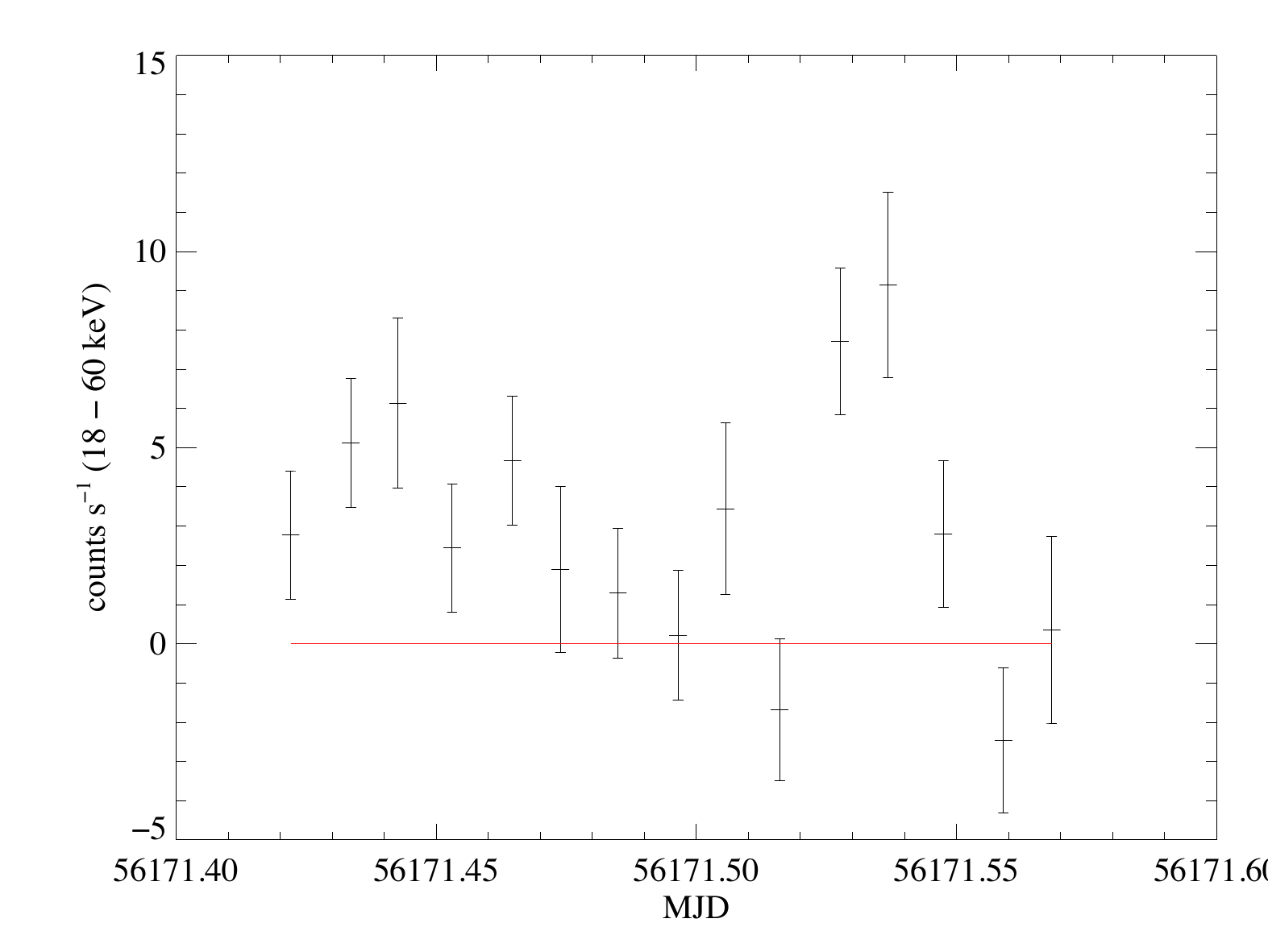}
	\caption{1000\,s binned 18 $-$ 60\,keV light curve of the low level activity observed during the first 5 ScWs of the recent \emph{INTEGRAL}/IBIS observations.}
\label{fig:IBIS1000seclc}
\end{figure}

The new \emph{INTEGRAL} observations were constrained in orbital phase to cover the eclipse region of IGR J16418$-$4532's orbit, covering $\phi$ $=$ 0.403 through 0.655 using the above ephemeris. The light curve is shown by the green points in Fig. \ref{fig:pfold} and it can be seen to cover the eclipse ingress and fully eclipsed region but ends just before the egress. IGR J16418$-$4532 shows a low, but rising level of activity in the first 4 ScWs before the eclipse which is then cutoff sharply in the fifth ScW at an orbital phase consistent with the eclipse ingress. The source is not detected during the subsequent ScWs performed during the eclipse. A mosaic image of the first 5 ScWs was produced in which IGR J16418$-$4532 was detected at a significance of 6.1$\sigma$. A spectrum was extracted from these ScWs and fit with a powerlaw in XSPEC version 12.7.1 \citep{1996ASPC..101...17A} with $\Gamma$\,$=$\,2.2$^{+1.1}_{-0.7}$ and $\hat{\chi}^{2}$\,$=$\,1.08(6\,\emph{dof}). The corresponding 18\,$-$\,60\,keV flux was 1.4\,$\times$\,10$^{-10}$\,erg\,cm$^{-2}$\,s$^{-1}$. A higher time resolution light curve with 1000\,s bins was also extracted from these ScWs. This light curve is shown in Fig. \ref{fig:IBIS1000seclc} and it again shows a low level of activity, followed by an increase in flux at MJD 56171.53 which then rapidly drops again. The peak emission occurs at MJD 56171.537$\pm$0.0058, relating to an orbital phase of $\phi$ $=$ 0.431. Whilst it is possible that the flux variation could be due to X-ray flaring the decrease in flux could also be resulting from eclipse ingress which would place a limit on the ingress time of $\sim$ 2000\,s ($\Delta\phi$ $=$ 0.006) and provides an accurate eclipse ingress ephemeride. Unfortunately the signal-to-noise was insufficient to probe this emission at higher time resolutions. 

IGR J16418$-$4532 was not detected in the soft 3\,$-$10\,keV X-ray band by the JEM-X instrument \citep{2003A&A...411L.231L} aboard \emph{INTEGRAL} during these recent observations. The 3$\sigma$ upper limit derived from the mosaic images of both JEM-X units for an exposure time of 14.5\,ks, the majority of which occurred whilst IGR J16418$-$4532 was in eclipse, is 3.4\,mCrab (3\,$-$\,10\,keV).

\begin{table*}
	\caption{Newly discovered outbursts of IGR J16418$-$4532 in the archival \emph{INTEGRAL}/IBIS data set. The orbital phase given relates to the ScW with the maximal count rate in each event. The peak flux is that of the same ScW in the 18\,-\,60\,keV band, calculated assuming a powerlaw spectrum with $\Gamma$\,$=$\,2.2 as observed in the activity at the start of the new \emph{INTEGRAL}/IBIS data set. Note: Fluxes denoted by a * are likely over estimates of the source flux due to large statistical uncertainties in the detected count rate.}
	\begin{center}
	\begin{tabular}{|c|c|c|c|c|c|}
	\hline
	\multicolumn{1}{|c|}{Significance} & \multicolumn{1}{|c|}{Start MJD} &  \multicolumn{1}{|c|}{End MJD} & \multicolumn{1}{|c|}{Duration (hours)} & \multicolumn{1}{|c|}{Orbital Phase $\phi$} & \multicolumn{1}{|c|}{Peak Flux (erg\,cm$^{-2}$\,s$^{-1}$)} \\ \hline
	6.3 & 52651.96 & 52652.21 & 5.8 & 0.14 & 2.3$\times$10$^{-9}$* \\ 
	11.8 & 52702.81 & 52703.61 & 19.4 & 0.71 & 2.2$\times$10$^{-10}$ \\
	7.0 & 52711.66 & 52712.08 & 10.3 & 0.12 & 2.8$\times$10$^{-10}$ \\
	7.3 & 52723.58 & 52724.23 & 15.6 & 0.37 & 1.7$\times$10$^{-9}$* \\
	7.3 & 52914.12 & 52914.59 & 11.3 & 0.29 & 4.8$\times$10$^{-10}$ \\
	4.6 & 53107.99 & 53108.54 & 13.2 & 0.05 & 2.1$\times$10$^{-9}$* \\
	7.1 & 53430.09 & 53430.16 & 1.7 & 0.21 & 2.5$\times$10$^{-10}$ \\
	6.6 & 53430.35 & 53430.38 & 0.7 & 0.27 & 2.7$\times$10$^{-10}$ \\
	5.2 & 53455.90 & 53455.91 & 0.2 & 0.10 & 3.1$\times$10$^{-10}$ \\
	8.3 & 54861.75 & 54861.81 & 1.7 & 0.13 & 4.6$\times$10$^{-10}$ \\
	7.2 & 54868.76 & 54868.83 & 1.7 & 0.99 & 5.4$\times$10$^{-10}$ \\
	5.7 & 55104.88 & 55104.92 & 1.2 & 0.15 & 3.2$\times$10$^{-10}$ \\
	4.5 & 55239.69 & 55239.72 & 0.5 & 0.20 & 5.8$\times$10$^{-10}$ \\
	6.6 & 55245.33 & 55245.71 & 9.1 & 0.72 & 2.5$\times$10$^{-10}$ \\
	5.6 & 55257.83 & 55257.91 & 1.9 & 0.07 & 2.8$\times$10$^{-10}$ \\
	8.6 & 55260.43 & 55261.11 & 16.3 & 0.77 & 2.2$\times$10$^{-10}$ \\
	6.2 & 55287.64 & 55287.66 & 0.5 & 0.03 & 5.1$\times$10$^{-10}$ \\
	5.9 & 55301.46 & 55301.60 & 3.1 & 0.73 & 2.7$\times$10$^{-10}$ \\
	5.9 & 55429.87 & 55430.02 & 3.6 & 0.07 & 2.1$\times$10$^{-10}$ \\
	\hline
	\end{tabular}
	\end{center}
	\label{tab:IBISbursts}
\end{table*}

%{\bf Note: I know some of the stuff from the INTEGRAL observations isn't the best quality but I think for political reasons it's good to have something in here about it. We don't want TAC members to read the paper and think you didn't even look at the new INTEGRAL data, you just used it to get the XMM time. The chances of this happening probably aren't that high but better safe than sorry for the sake of future proposals :-P.}

\label{sect:IDR}

\section{\emph{XMM-Newton} data analysis and results}

Two separate \emph{XMM-Newton}/EPIC (\citealt{2001A&A...365L...1J}, \citealt{2001A&A...365L..27T}, \citealt{2001A&A...365L..18S}) observations were performed quasi-simultaneously with \emph{INTEGRAL} to achieve sensitive soft X-ray coverage over a wider region of orbital phase. The first observation was performed between UTC 21:12:25 2012-09-01 and 02:53:22 2012-09-02 for an exposure of $\sim$20\,ks. This observation covered the orbital phase range $\phi$ $=$ 0.525 $-$ 0.587 and occurred whilst the NS was fully eclipsed by the supergiant. The second observation was performed after the NS had egressed from the eclipse, between UTC 14:16:33 and 19:12:03 2012-09-02, for an exposure of $\sim$14\,ks and covered the orbital phase range $\phi$ $=$ 0.715 $-$ 0.758. The phase location of these observations is shown in Fig. \ref{fig:pfold}. In both observations the EPIC-MOS and EPIC-pn detectors were operating in large window mode.

Data from both the EPIC-MOS and EPIC-pn detectors were analysed from each observation using SAS v12.0.1 \citep{2004ASPC..314..759G} and the most recent instrument calibration files. The data sets were checked for regions of high particle background following the method outlined in the \emph{XMM-Newton} SAS data analysis threads\footnote{http://xmm.esac.esa.int/sas/current/documentation/threads/} with no regions of high particle background identified in either the MOS or pn 10\,$-$\,12\,keV light curves using cut-offs of 0.35 and 0.4 counts\,s$^{-1}$ respectively. The EREGIONANALYSE tool was used to define the optimal extraction region for all light curve and spectral generation procedures (note: all subsequent references to optimal extraction regions were defined using this method). The data sets were evaluated for the presence of photon pile-up using the EPATPLOT tool and by comparing the shape of spectra extracted from both circular and annular extraction regions. The observations were found to be unaffected by pile-up apart from the final few ks of the second data set, during which the shape of the extracted spectra was observed to vary slightly when circular and annular extraction regions were used. As such the standard annular extraction region method was utilised to remove the effects of pile-up from the extracted X-ray spectra during these times. The specific extraction region shapes and sizes used for both data sets are outlined in Sections \ref{sect:MEO} and \ref{sect:PEO}. All spectra reported in this work were extracted following the standard procedures and the SAS tools RMFGEN and ARFGEN were used to extracted the necessary response files for each spectrum. Spectra were again fit using XSPEC version 12.7.1 with uncertainties quoted at the 90\% confidence level throughout and the elemental abundances set to those of \citet{2000ApJ...542..914W}. Due to the very different nature of the detected emission in the two observations, the detailed analysis of each is reported separately in Sections \ref{sect:MEO} and \ref{sect:PEO} below.

\label{sect:XDR}

\subsection{Mid-eclipse Observation}

\begin{figure}
	\includegraphics[width=0.45\textwidth]{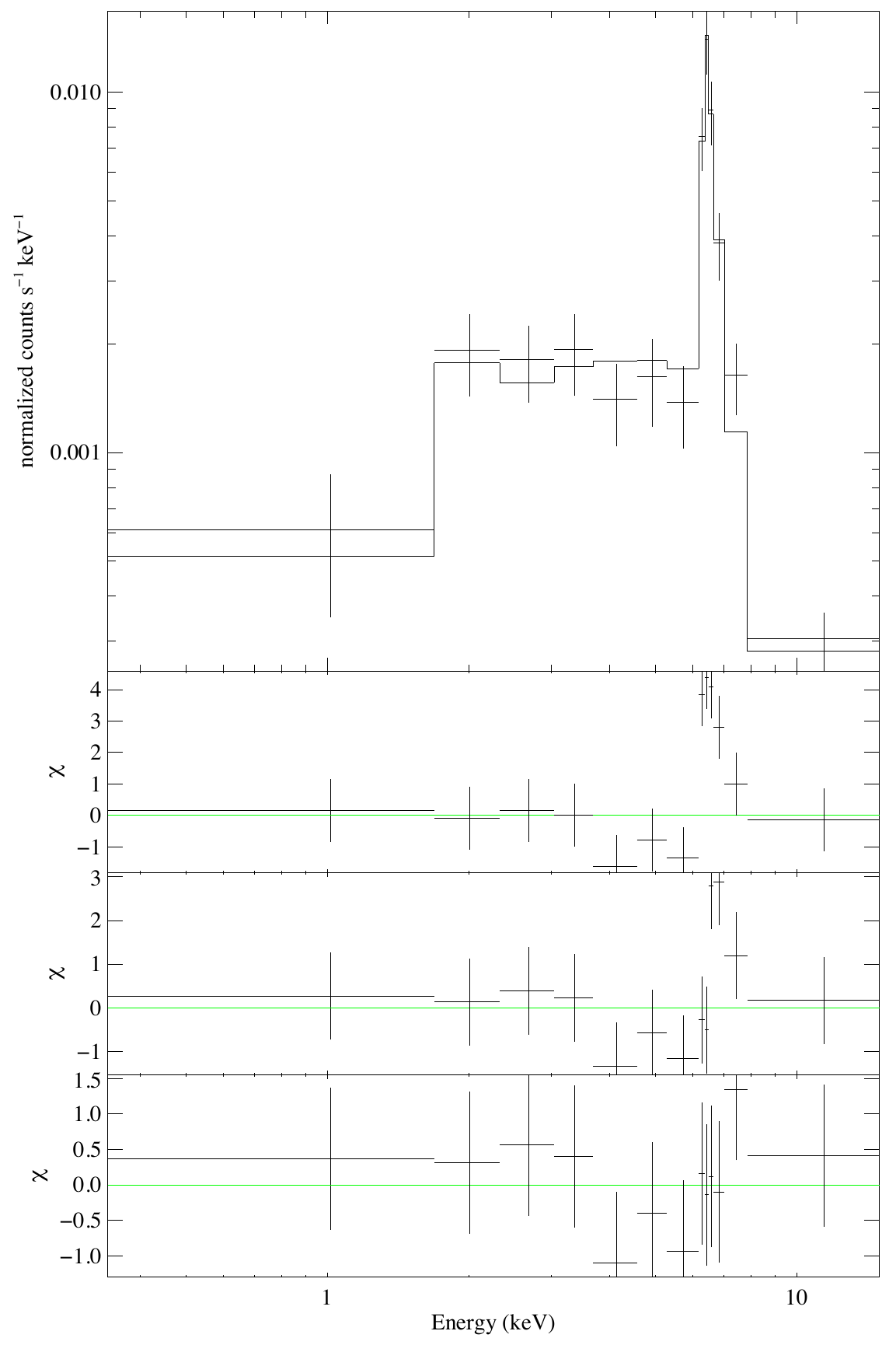}
	\caption{Top panel: EPIC-pn best fit spectrum of IGR J16418$-$4532 during the MEO. The spectrum is best fit with an absorbed powerlaw with two intrinsically narrow Gaussian emission lines at energies of 6.40\,$^{+0.03}_{-0.04}$ and 6.65\,$^{+0.06}_{-0.05}$\,keV respectively. Lower panels: The three lower panels show the residuals of the fit when using an absorbed powerlaw, an absorbed powerlaw with one intrinsically narrow Gaussian and the best fit absorbed powerlaw with two intrinsically narrow Gaussians.}
\label{fig:OBS1spec}
\end{figure}

Emission from IGR J16418$-$4532 was detected at a low level in the mid-eclipse observation (MEO) which, given its location in orbital phase, is assumed to originate from reprocessing of the the NS X-ray flux by the stellar wind of the supergiant. There was insufficient signal to produce meaningful light curves and spectra in the MOS detectors so they will not be considered further. Similarly in the pn detector only a low signal-to-noise light curve could be extracted in a broad (0.2 $-$ 10\,keV) energy band. The light curve did not provide any detailed temporal information, however a spectrum was accumulated from the full exposure of the pn detector using an optimal circular extraction region of radius 14$^{\prime\prime}$. The background subtracted pn spectrum of the MEO, binned to a minimum of 25 counts per bin, is shown in the top panel of Fig. \ref{fig:OBS1spec}. It can be seen that the spectrum has a flat shape with a large emission feature consistent with a 6.4\,keV iron-K$\alpha$ line. 

To further characterise the spectral shape of the pn spectrum it was first fit with a simple absorbed powerlaw continuum with an additional Gaussian component (PHABS(POWERLAW + GAUSSIAN) in XSPEC). The absorption was fixed to the Galactic value in the direction of IGR J16418$-$4532, 1.88$\times$10$^{22}$ cm$^{-2}$ \citep{1990ARA&A..28..215D}, as in this observation the powerlaw is simply parameterising a more complicated reprocessing mechanism and is unlikely to be describing the intrinsic, locally absorbed continuum emission. The power law index, Gaussian line energy and line width were left as free parameters. The spectrum was well fit by this model with $\hat{\chi}^{2}$ $=$ 1.02 (8 \emph{dof}), however the Gaussian line width $\sigma$$_{E}$ $=$ 0.15\,keV (EW\,$=$3.1\,keV) corresponds to an electron plasma temperature of $\sim$3$\times$10$^{6}$\,K which is well in excess of that expected for a supergiant atmosphere. This unphysical fit is likely resulting from the low signal-to-noise and coarsely sampled spectrum combined with the possible influence of the 7.1\,keV iron absorption edge located within the emission line region of the spectrum.  

\begin{table}
	\caption{IGR J16418$-$4532 MEO best fit EPIC-pn spectral parameters and uncertainties. The spectrum was fit using the model PHABS(POWERLAW + GAUSS + GAUSS) with n$_{H}$ fixed and $\hat{\chi}^{2}$ $=$ 0.72 (7 \emph{dof}). The flux is quoted in the 0.5$-$10\,keV band}
	\begin{center}
	\begin{tabular}{|c|c|c|}
	\hline
	\multicolumn{1}{|c|}{Parameter} & \multicolumn{1}{|c|}{Value} & \multicolumn{1}{|c|}{Unit} \\ \hline
	n$_{H}$ & 1.88$\times$10$^{22}$ & cm$^{-2}$ \\ \smallskip
	$\Gamma$ & 0.29\,$^{+0.3}_{-0.3}$ \\ \smallskip
	E$_{1}$ centroid & 6.40\,$^{+0.03}_{-0.04}$ & keV \\ \smallskip
	E$_{1}$ EW & 0.63\,$^{+0.28}_{-0.16}$ & keV \\ \smallskip
	E$_{1}$ intensity & 4.9$\times$10$^{-6}$ & ph\,cm$^{-2}$\,s$^{-1}$ \\ \smallskip
	E$_{2}$ centroid & 6.65\,$^{+0.06}_{-0.05}$ & keV \\ \smallskip
	E$_{2}$ EW & 0.33\,$^{+0.17}_{-0.15}$ & keV \\ \smallskip
	E$_{2}$ intensity & 3.3$\times$10$^{-6}$ & ph\,cm$^{-2}$\,s$^{-1}$ \\ \smallskip
	flux & (2.80\,$^{+0.34}_{-0.42}$)\,$\times$10$^{-13}$ & erg\,cm$^{-2}$\,s$^{-1}$ \\ \hline
	\end{tabular}
	\end{center}
	\label{tab:OBS1param}
\end{table}

To obtain a more physical characterisation of the iron line profile the width of the Gaussian line was fixed at 0 such that the line is intrinsically narrow and broadened only by the EPIC-pn response. Using a single Gaussian with the centroid energy as a free parameter provided a poor fit to the data ($\hat{\chi}^{2}$ $=$ 2.04 (9 \emph{dof})) with large residuals in the region of the emission line. A second intrinsically narrow Gaussian was added at higher energy to account for emission from ionised iron and the spectrum fit with the centroid energy left as a free parameter. The resulting fit was of good quality with $\hat{\chi}^{2}$ $=$ 0.72 (7 \emph{dof}) and the line centroid energies E$_{1}$ $=$ 6.40\,$^{+0.03}_{-0.04}$ and E$_{2}$ $=$ 6.65\,$^{+0.06}_{-0.05}$\,keV, consistent with both neutral 6.4\,keV and ionised 6.67\,keV Fe-K$\alpha$ emission. The equivalent widths of the 6.40 and 6.65\,keV lines were 0.63 and 0.33\,keV respectively showing that the neutral emission is approximately twice the strength of the ionised emission. The 0.5 $-$ 10\,keV detected flux was (2.80\,$^{+0.34}_{-0.42}$)$\times$10$^{-13}$\,erg cm$^{-2}$ s$^{-1}$. The best fit parameters resulting from this model are outlined in full in Table \ref{tab:OBS1param} with the best fit model and residuals shown in Fig. \ref{fig:OBS1spec}, where the lower panels illustrate the improvement in the residuals as the Gaussian lines are added to the model.

%For completeness a third Gaussian with centroid energy 6.97\,keV was also added to the model to account for Fe-K$\alpha$ emission from a higher ionisation state. Whilst this model also produced an acceptable fit with $\hat{\chi}^{2}$ $=$ 0.998 (5 \emph{dof}), the residuals about the Fe emission feature were not improved by the inclusion of this additional spectral component. Hence there was insufficient justification to include the extra spectral component and this model was not considered further. 

%{\bf Is there a better/more physical model to use for the continuum fit? Is there a method of statistically assessing the need for the extra Gaussians given that F-tests shouldn't be used on Gaussian lines?}
 
\label{sect:MEO}

\subsection{Post-eclipse Observation}

\begin{figure*}
	\includegraphics[width=\textwidth,height=0.4\textheight]{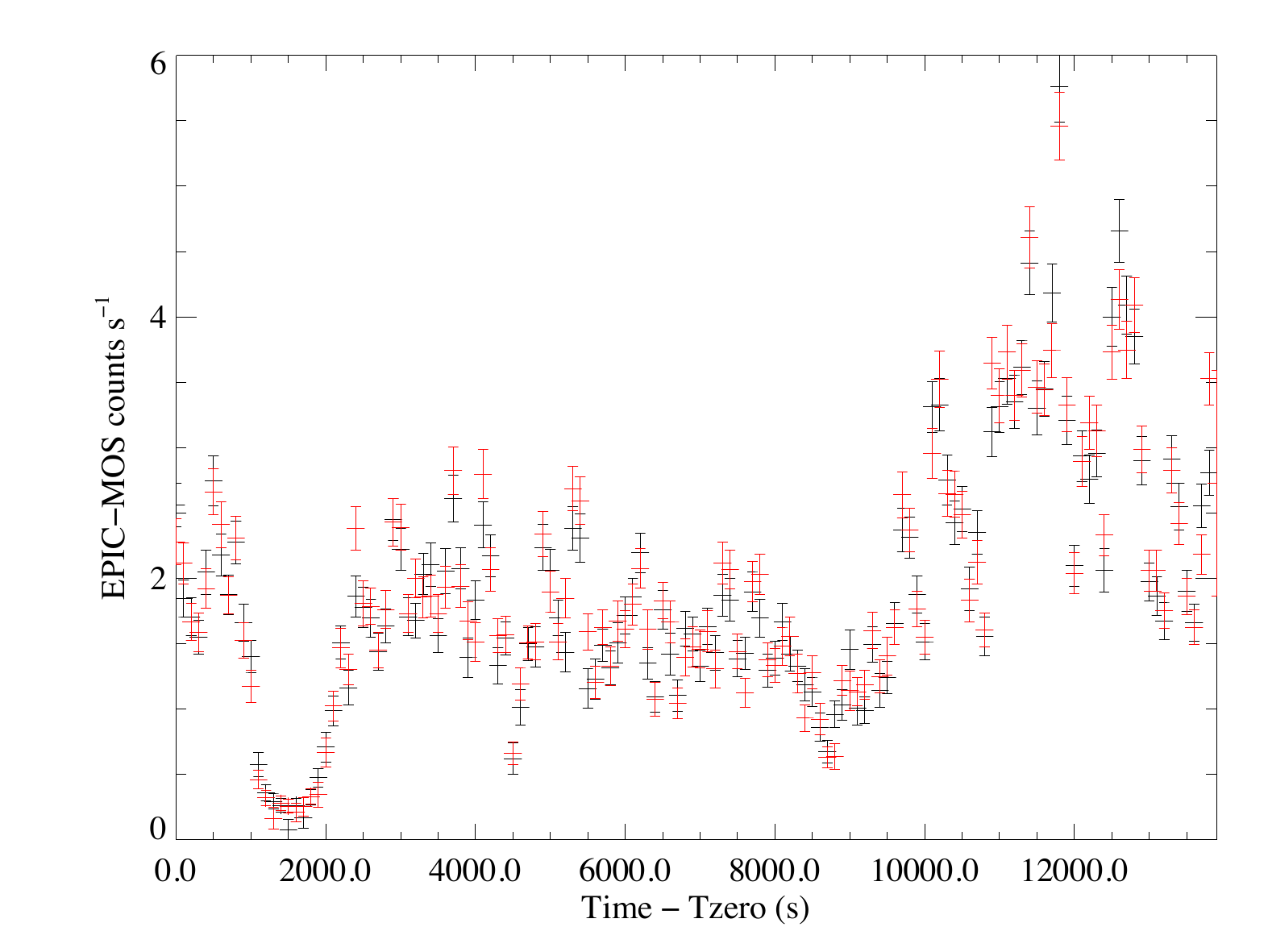}
	\caption{EPIC-MOS1 (black) and MOS2 (red) 100\,s binned light curve of IGR J16418$-$4514 in the 0.2 $-$ 10\,keV band}
	\label{fig:OBS2MOSlcs}
\end{figure*}

\begin{figure}
	\includegraphics[width=0.5\textwidth]{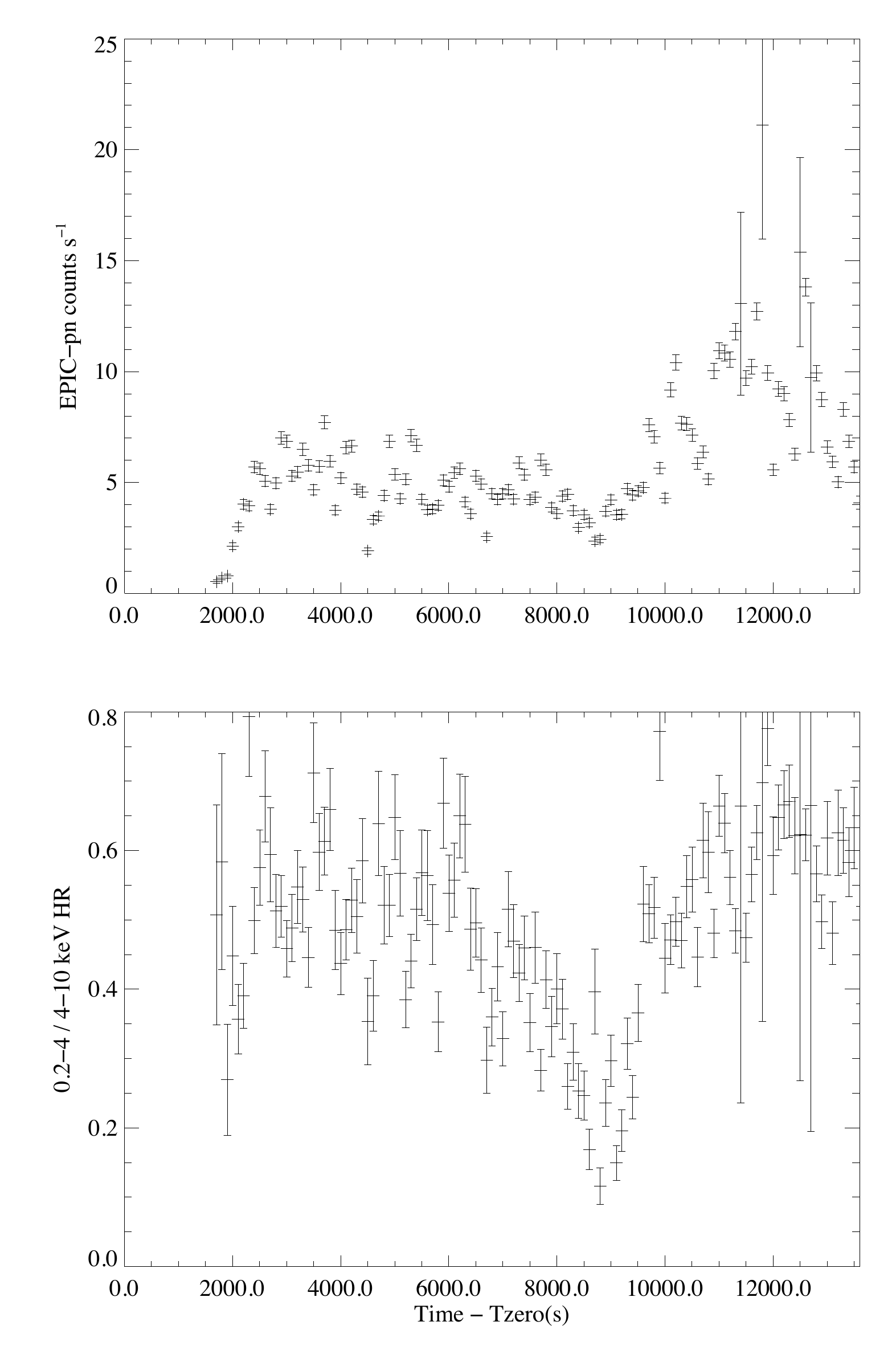}
	\caption{Top: EPIC-pn 100\,s binned light curve of IGR J16418$-$4532 in the 0.2$-$10\,keV band. Bottom: EPIC-pn 0.2$-$4 / 4$-$10\,keV hardness ratio light curve at the same resolution where lower values indicate a hardening of the detected flux}
	\label{fig:OBS2PNHRlc}
\end{figure}

During the post-eclipse observation (PEO) IGR J16418$-$4532 was detected with good signal-to-noise. Figure \ref{fig:OBS2MOSlcs} shows the 0.2 $-$ 10\,keV EPIC-MOS1 (black) and MOS2 (red) light curves, extracted from optimally defined circular regions of radius 97$^{\prime\prime}$ and 98$^{\prime\prime}$ respectively, with time binning of 100\,s. T$_{zero}$ is defined as the first time stamp in the MOS1 light curve and takes the value T$_{zero}$\,$=$\,MJD\,56172.596. The light curves reveal a range of activity including a steady emission at $\sim$ 2 counts s$^{-1}$, a rapid `dipping' feature between t $\sim$ 800 and 2500\,s and flaring behaviour at times of t $>$ 10,000\,s. There is also activity on short, hundreds of seconds timescales throughout the exposure, some of which may be due to the $\sim$1200\,s pulse period. This activity was also observed in the EPIC-pn detector, the 100\,s binned light curve of which is shown in the top panel of Fig. \ref{fig:OBS2PNHRlc}. The pn light curve was extracted from an optimal circular region of radius 68$^{\prime\prime}$. Due to additional observational overheads the pn detector did not start exposing until t $=$ 1700\,s, resulting in a lack of coverage during the majority of the `dip' region observed by the MOS detectors. To account for the effects of pile-up at times greater than t\,$=$\,10,000\,s, annular extraction regions with inner radii of 7.5$^{\prime\prime}$ and 10$^{\prime\prime}$ for the MOS and pn detectors respectively were used to extract all data products. 

The maximal dynamic range inferred from the 100\,s binned broad band 0.2$-$10\,keV light curves is 75 (MOS1) indicating a large variation in the flux detected during the relatively short exposure. 100\,s binned light curves were also extracted in restricted 0.2$-$4 and 4$-$10\,keV energy bands and the 0.2$-$4 / 4$-$10\,keV hardness ratio (HR) calculated for each detector. The EPIC-pn HR is shown in the lower panel of Fig. \ref{fig:OBS2PNHRlc} and it is seen to be composed of features on varying time scales. The most striking feature is the systematic hardening of the detected emission between t $=$ 6000 and 10,000\,s and the strong correlation between the HR and detected source flux minima. The end of this hardening is also temporally coincident with the onset of the flaring behaviour observed from t\,$=$\,10,000\,s onwards. Variation is also observed on timescales of a few 100\,s throughout the observation, regardless of flux level, and there is evidence of systematic softening of the detected emission as the source exits the `dip' region of the light curve. HR light curves were also generated for the MOS detectors that showed the same evolution through the observation, albeit with less clarity due to the larger count rate errors, but could not provide additional diagnostics during the `dip' region as a result of the limited photon statistics at this time.

\label{sect:PEO}

\subsubsection{Spectral Analysis}

The lower panel of Fig. \ref{fig:OBS2PNHRlc} illustrates the level and speed of variation in the emission of IGR J16418$-$4532 during the PEO through the evolution of the hardness ratio, implying fast variations in the detected source spectrum. In order to track these variations we extracted spectra accumulated from bins of length 400\,s in steps of 200\,s through the length of the exposure. An image was extracted for each time bin and the optimal extraction region for that region defined by the combination of EREGIONANALYSE and the circle/annulus shape constraints described above. Spectra were extracted, and binned to a minimum of 15 counts per bin, from both the MOS and pn detectors at times greater than t $=$ 1700\,s and the MOS detectors only prior to this. Such a systematic approach was adopted to remove possible biases introduced by the subjective selection of `similar' regions while the extraction bin size of 400\,s was chosen as a compromise between generating spectra of sufficient signal-to-noise to perform accurate fitting and reducing the amount of HR variation within each bin. The spectra were binned at 15 counts per bin to ensure a sufficient number of spectral bins to allow model fitting were present even in the spectra accumulated from the time periods of lowest flux. Step sizes of 200\,s were chosen so that the evolution of spectral parameters could be followed at sufficient resolution but the data would only be over sampled by a factor of 2. 

%With this in mind it is uninformative to accumulate spectra from large sections of the exposure as the mixing of spectral states will likely prevent the accurate fitting of single spectral models and/or lead to an incorrect characterisation of the spectral shape. 

%\begin{figure*}
%	\begin{center}
%	\includegraphics[width=\textwidth]{IGRJ16418_HTR_spectral_parameters_400secbins_PL_winset}
%	\end{center}
%	\caption{Temporal evolution of the best fit spectral parameters of an absorbed powerlaw to the 400\,s binned spectra. Clock wise from top left: EPIC-MOS2 0.2\,$-$\,10\,keV light curve; Powerlaw index evolution; n$_{H}$ evolution; inferred 0.5\,$-$\,10\,keV flux where the diamond points represent fluxes derived from the spectra accumulated from annular extraction regions. Inset: Reduced $\chi^{2}$ values for the fit to each spectrum. {\bf I will make these into 4 by 1 plots for the final version}}
%	\label{fig:OBS2specparam}
%\end{figure*}

\begin{figure}
	\begin{center}
	\includegraphics[width=0.45\textwidth]{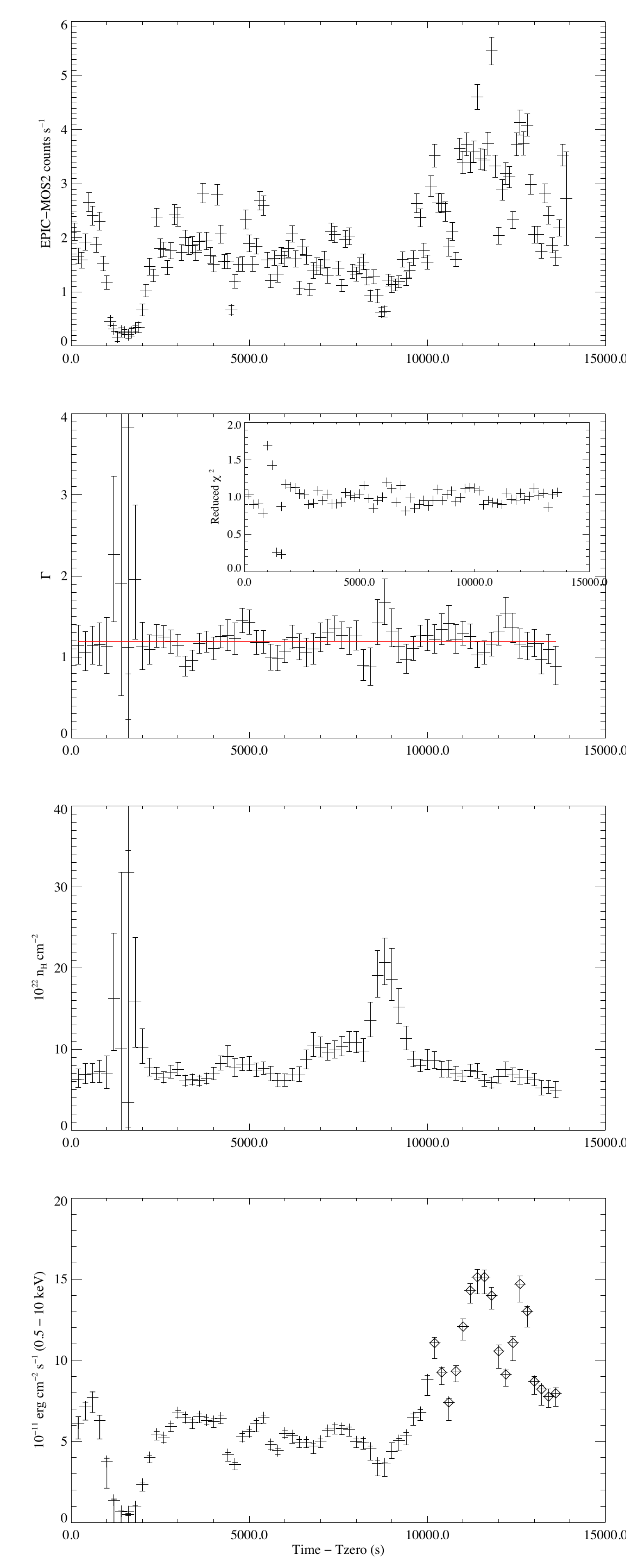}
	\end{center}
	\caption{Temporal evolution of the best fit spectral parameters of an absorbed powerlaw to the 400\,s binned spectra. Top to bottom: EPIC-MOS2 0.2\,$-$\,10\,keV light curve; Powerlaw index evolution; n$_{H}$ evolution; inferred 0.5\,$-$\,10\,keV absorbed flux where the diamond points represent fluxes derived from the spectra accumulated from annular extraction regions. Inset: Reduced $\chi^{2}$ values for the fit to each spectrum.}
	\label{fig:OBS2specparam}
\end{figure}

Inspection of the spectra showed no evidence of Fe emission lines so the spectra were fit with simple absorbed models to characterise their shapes. Absorbed powerlaws (PHABS(POWERLAW) in XSPEC) were simultaneously fit to the spectra from all available detectors in each time bin in the 0.5\,$-$\,15\,keV energy range with the absorption, powerlaw index and normalisation left as free parameters. The evolution of the spectral parameters, uncertainties and derived fluxes across the PEO are shown in Fig. \ref{fig:OBS2specparam}. The inset panel in Fig. \ref{fig:OBS2specparam} shows the $\hat{\chi}^{2}$ value of the fit to each bin throughout the observation. For the majority of the exposure time the $\hat{\chi}^{2}$ values are distributed about 1.0 showing a good fit to the spectra. However bins falling within the `dip' region of the observation, in particular those occurring during the deepest part of the `dip', gave unacceptable $\hat{\chi}^{2}$ values, indicating that an absorbed powerlaw is a poor description of the emission at these times. The spectra of this region are considered separately and in more detail below. For the rest of the observation it can be seen that when both n$_{H}$ and $\Gamma$ are left as free parameters there is a level of variation in both parameters. However a Chi-Squared test to the mean value of each parameter, where both the mean and Chi-squared statistic have been calculated with the poorly fitted bins omitted, gave a  $\hat{\chi}^{2}$ value of 0.65 (67 \emph{dof}) and 3.52 (67 \emph{dof}) for $\Gamma$ and n$_{H}$ respectively. Hence the $\Gamma$ parameter is consistent with being constant throughout the observation whilst the absorption shows excess variation. The average $\Gamma$ value is 1.195 and is over plotted as the red line in the second panel of Fig. \ref{fig:OBS2specparam}.   

\begin{figure}
	\begin{center}
	\includegraphics[width=0.45\textwidth]{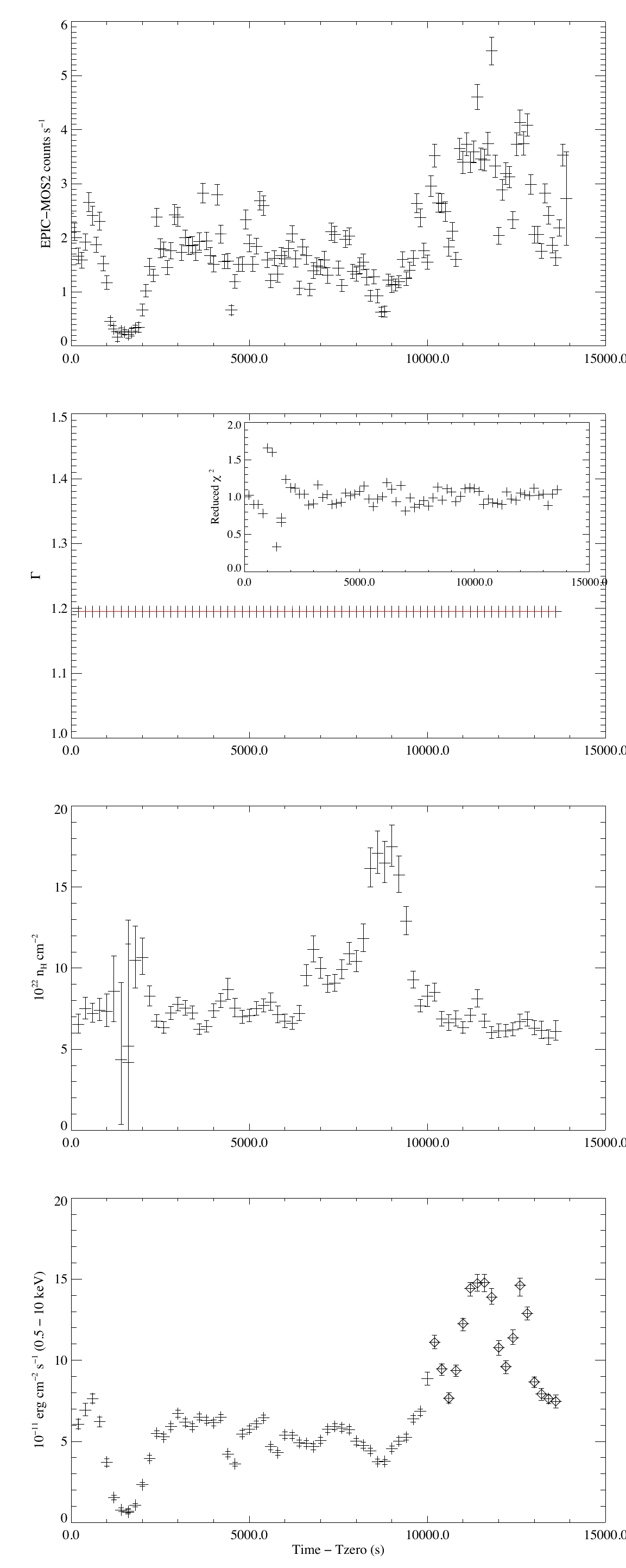}
	\end{center}
	\caption{Temporal evolution of the best fit spectral parameters of an absorbed powerlaw with a fixed $\Gamma$\,$=$\,1.195. The panels show the same plots as Fig. \ref{fig:OBS2specparam}.}
	\label{fig:OBS2specparamfixedGamma}
\end{figure}

A second set of fits was then performed with $\Gamma$ fixed at the value of 1.195, whilst the normalisation and n$_{H}$ were left as free parameters. The evolution of the spectral parameters and the goodness-of-fit in this case are shown in Fig. \ref{fig:OBS2specparamfixedGamma}. A similar, better constrained n$_{H}$ evolution as that seen in Fig. \ref{fig:OBS2specparam} is again observed when fitting a powerlaw of fixed $\Gamma$ with this model providing an acceptable fit for all bins apart from those occurring during the `dip' feature. The n$_{H}$ evolution shows that there is a constant, high level of absorption of n$_{H}$\,$\sim$7$\times$10$^{22}$\,cm$^{_2}$ that is intrinsic to the IGR J16418$-$4514 system (the Galactic value in the direction of the source is 1.88$\times$10$^{22}$ cm$^{-2}$). The most striking feature of the evolution is the large, structured increase in n$_{H}$ between t\,$=$\,6500 and 9500\,s which is co-incident with the `Pre-flare hardening' feature observed in the HR. In particular the peak n$_{H}$ is again temporally coincident with the HR and detected flux minima. Figure \ref{fig:OBS2nHunabsf} shows an expanded view of the n$_{H}$ evolution overlaid with the unabsorbed flux from each bin, calculated assuming the fixed powerlaw of $\Gamma$\,$=$\,1.195. It is seen that removing the effects of absorption brings the flux back up to the level seen before the $n_{H}$ increase but it is not until after the n$_{H}$ has returned to its base level that the large increase in flux starts to occur. Consideration of the physical interpretation of this behaviour is given in Section \ref{sect:DISC}. Outside of the large increase the n$_{H}$ also shows some level of variation that occurs on the same time scale as the NS pulsation ($\sim$\,1200\,s). This variation is more pronounced in the spectral fits performed with a fixed $\Gamma$, adding to the justification for fixing the photon index as this produces spectral variations that may be linked to a physical origin whereas fixing the n$_{H}$ and allowing $\Gamma$ to vary only produces fits with very poor $\hat{\chi}^{2}$ values.

%\begin{figure*}
%	\begin{center}
%	\includegraphics[width=\textwidth]{IGRJ16418_HTR_spectral_parameters_400secbins_PL_fixedGamma_winset}
%	\end{center}
%	\caption{Temporal evolution of the best fit spectral parameters of an absorbed powerlaw with a fixed $\Gamma$\,$=$\,1.195. The panels show the same plots as Fig. \ref{fig:OBS2specparam}.}
%	\label{fig:OBS2specparamfixedGamma}
%\end{figure*}

The time resolved spectra were also fit with the more complicated `Partial Covering' absorption model (PCFABS in XSPEC) in an attempt to better understand the evolution of the absorbing material during the PEO. However it was seen that the fits were insensitive to the additional parameters, with the covering fraction tending towards 100\% for all spectra while the total n$_{H}$ (the combination of a component fixed to the galactic value and a free varying, partial covering component) was consistent with the best fit value for the simple absorbed powerlaw model. Hence no additional information could be drawn from this model. The time resolved spectral analysis was performed with different bin durations, step sizes and over sampling factors and the evolution was observed to follow a consistent path regardless of the extraction used. As a final test a blackbody continuum was used in place of the powerlaw model. A consistent n$_{H}$ and flux evolution, with a similar distribution of $\hat{\chi^{2}}$ values, was observed throughout the observation producing further evidence that the evolution is explained by a relatively constant continuum modified by a varying photoelectric absorption. For the remainder of this work we refer only to the simple absorbed powerlaw fits when discussing the spectral evolution. 

To characterise the emission during the dip region of the light curve a separate EPIC-pn spectrum was accumulated encompassing all available EPIC-pn data of the region, namely between t\,$=$\,1678\,s and t\,$=$\,2000\,s for a total exposure of 322\,s. The spectrum showed no indication of Fe emission lines and was fit with a simple absorbed powerlaw where $\Gamma$ was fixed to 1.195 (as above) and the best fit n$_{H}$\,$=$\,(11.1$^{+2.8}_{-2.1}$)\,$\times$10$^{22}$\,cm$^{-2}$. The $\hat{\chi^{2}}$\,$=$\,0.72(15 \emph{dof}) and the corresponding 0.5\,$-$10\,keV unabsorbed flux was 2.0$\times$10$^{-11}$\,erg\,cm$^{-2}$\,s$^{-1}$. When compared to the maximum, unabsorbed flux detected later in the observation (27.2$\times$10$^{-11}$\,erg\,cm$^{-2}$\,s$^{-1}$ (0.5\,$-$\,10\,keV)) a soft X-ray dynamic range of 14 is observed during the PEO.

\label{sect:PEOspec}

\subsubsection{Timing Analysis}

To test for signatures of the known 1212$\pm$6\,s NS spin period \citep{2012MNRAS.420..554S} the 0.2$-$10\,keV light curves from each instrument were re-extracted with a variety of bin sizes. Here we report on the results from light curves with a binning of 10\,s but point out that the results were consistent in the analysis of each different binning. For the purposes of temporal analysis the `dip' region of each light curve was removed as it occurs over a similar timescale to the known pulsation and could therefore generate significant power at frequencies similar to that of the NS pulse period, despite it not necessarily being of a rotational nature. The Lomb-Scargle technique was again applied to the light curves and strong signals detected at $\sim$403\,s in all detectors. From here we consider only the MOS2 light curve as it has a longer baseline than that of the pn and the results are consistent with those from MOS1. The uncertainty on the peak was calculated using the method outlined in Section \ref{sect:IDR} and the peak period found to be 403.04$\pm$0.14\,s. This is consistent with being the third harmonic of the known pulse period with the fundamental frequency relating to a period of 1209.12$\pm$0.42\,s. This spin period is consistent with the previously reported values, summarised in \citet{2012MNRAS.420..554S}, and with the addition of this detection there is still no significant detection of spin period evolution in IGR J16418$-$4532. 

The upper and middle panels of Fig. \ref{fig:pulseHR} shows the MOS2 light curve folded on the fundamental period of 1209.12\,s in the 0.2$-$4\,keV and 4$-$10\,keV bands respectively. A similar pulse profile, with one sharp narrow peak and one equally high but broader peak, is detected in both energy bands with the 0.2$-$4\,keV displaying a pulse fraction of 25.0$\pm$3.6\% and the 4$-$10\,keV band 37.2$\pm$2.7\%. The pulse fraction of the folded profile in the broad 0.2$-$10\,keV band was 31.2$\pm$2.1\%. A double peaked profile is consistent with the 2011 \emph{XMM-Newton} observations of \citet{2012MNRAS.420..554S} and shows a continued departure from the 2004 \emph{XMM-Newton} observations of \citet{2006A&A...453..133W} which showed a pulse profile with a single, broad peak \citep{2012MNRAS.420..554S}. The double peaked profile with uneven pulse widths could be the driving factor behind the detection of the third harmonic, as opposed to the fundamental frequency in these observations. The lower panel of Fig. \ref{fig:pulseHR} shows the hardness ratio as a function of pulse phase and it is seen that there is no systematic variation of the hardness with the progression of the pulsation. 

\label{sect:PEOtiming}

\section{Discussion}

%\begin{figure}
%	\includegraphics[width=0.5\textwidth]{IGRJ16418_Roche_limits_martinst6}
%	\caption{Ratio of L1 point separation to stellar radii for Galactic O-type supergiant stars as a function of stellar mass.}
%	\label{fig:roche} 
%\end{figure}

\subsection{Stellar and Orbital Parameters}

IGR J16418$-$4532 is an intermediate SFXT that has shown some peculiarities in its emission history. Here we have presented new combined \emph{INTEGRAL} and \emph{XMM-Newton} observations of the source that targeted the eclipse region of IGR J16418$-$4532s orbit. The well determined orbital period of 3.73886$\pm$0.00028\,days \citep{2011ApJS..196....6L} allows us to place constraints on the orbital geometry and nature of the companion star. The flat, unstructured phase-folded light curve and orbital phase distribution of outbursts (outside of the X-ray eclipse) detected by \emph{INTEGRAL}, as shown in Fig. \ref{fig:pfold} and \ref{fig:burstphase} respectively, indicate that the orbit is likely circular ($e \sim 0$) due to the lack of coherent modulation of the hard X-ray flux across the orbital phase. A similar flat orbital profile is also seen in IGR J16479$-$4514 which has a short 3.32\,day eclipsing orbit \citep{2009MNRAS.397L..11J}. Conversely coherent orbital phase modulation has been observed in other SFXT systems which show evidence of a significant eccentricity in their orbit, such as a strongly peaked orbital emission profile with outbursts clustered around times of presumed periastron, i.e. IGR J17354$-$3255 \citep{2011MNRAS.417..573S}. Additional dynamical considerations also suggest IGR J16418$-$4532 should possess a circular geometry as with such a short orbital period the system would be expected to circularise relatively early in the lifetime of the X-ray binary. 

\begin{figure}
	\includegraphics[width=0.5\textwidth]{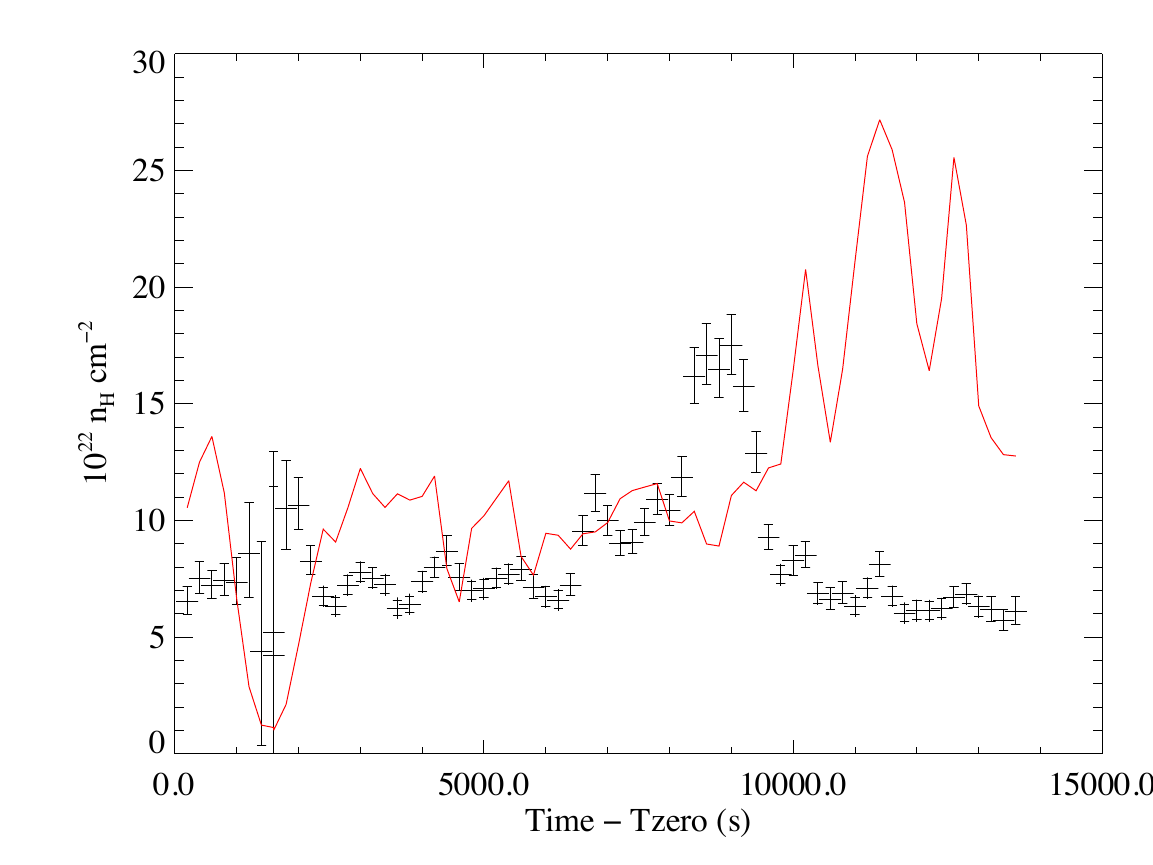}
	\caption{n$_{H}$ evolution through the PEO for an absorbed powerlaw with a fixed $\Gamma$\,$=$\,1.195 overlaid with the calculated unabsorbed 0.5$-$10\,keV flux in units of 10$^{-11}$ erg cm$^{-2}$ s$^{-1}$}
	\label{fig:OBS2nHunabsf}
\end{figure} 

\begin{figure}
	\includegraphics[width=0.5\textwidth]{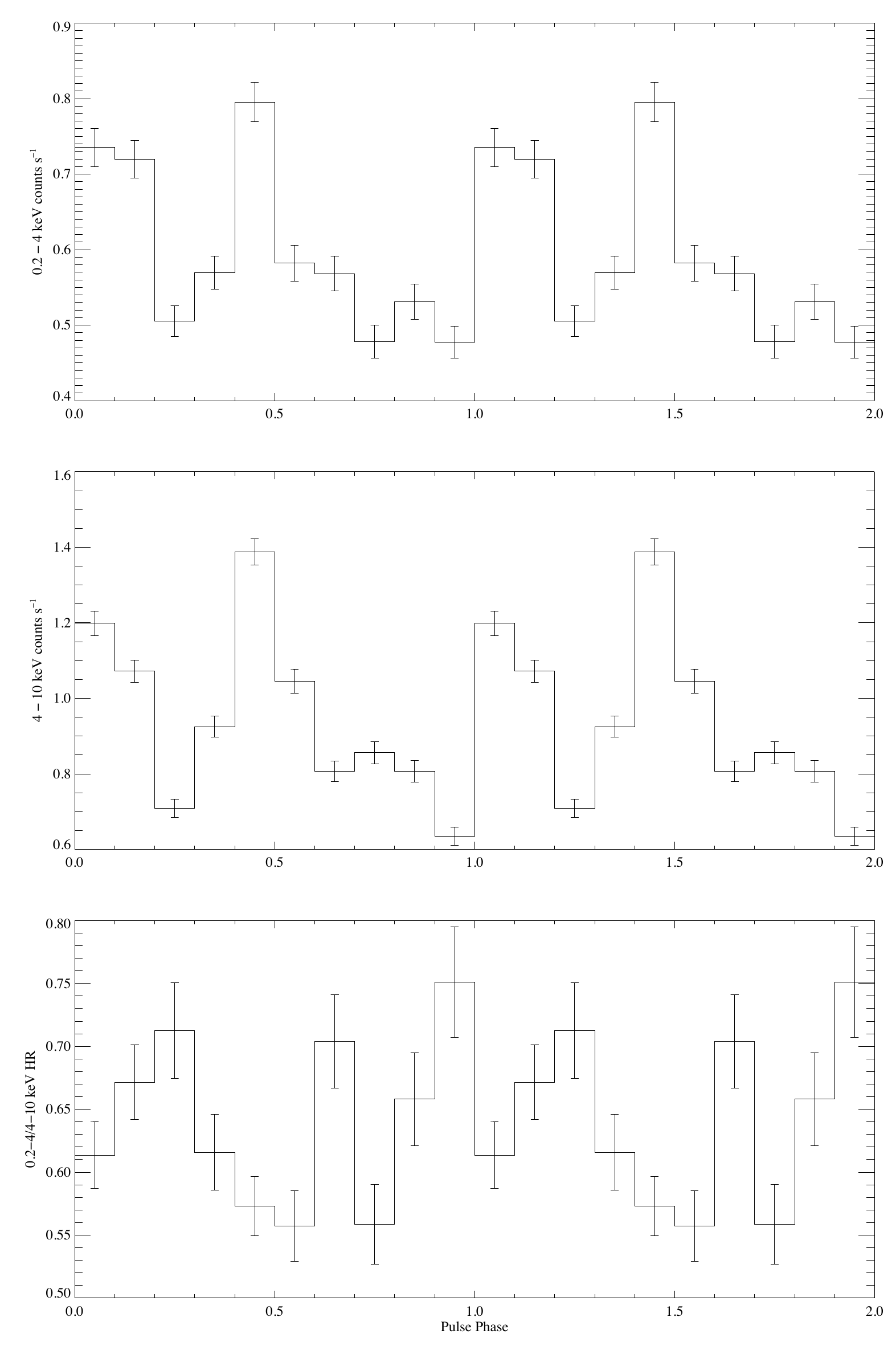}
	\caption{Pulse phase folded profiles using the pulse period of 1209.12\,s where the zero phase is arbitrary. Top: 0.2$-$4\,keV folded profile. Middle: 4$-$10\,keV folded profile. Bottom: HR of folded profile.}
	\label{fig:pulseHR}
\end{figure} 

The stellar parameters of the supergiant companion in IGR J16418$-$4532 have very few dynamic or spectroscopic constraints however. By combining the spectral classification (Coleiro et al. 2013, private communication), the IR SED fit parameters \citep{2008A&A...484..783C}, the well determined orbital period \citep{2011ApJS..196....6L} and the the L1 point separation equations of \citet{1983ApJ...268..368E} we can attempt to place some constraints on the orbital parameters through the condition that full RLO does not occur in IGR J16418$-$4532. Throughout these considerations we assume a neutron star in a circular orbit with a mass of 1.4\,M$_{\odot}$, a radius of 10\,km and a source distance of 13\,kpc \citep{2008A&A...484..783C}. We utilise the $R_{Sg}/D_{Sg}$ SED parameter of \citet{2008A&A...484..783C} to calculate the range of possible stellar radii of the companion star, the central value of which (3.77\,$\times$\,10$^{-11}$) gives a stellar radius of 21.7\,R$_{\odot}$. However, calculations of the L1 point separation for a range of companion star masses indicate that RLO would be occurring for all companion masses of less than 50\,M$_{\odot}$. Alternatively the lower limit of the $R_{Sg}/D_{Sg}$ parameter (2.64\,$\times$10$^{-11}$, 90\% confidence) gives a stellar radius of 15.2\,R$_{\odot}$ and RLO is only initiated up to a stellar mass of 20\,M$_{\odot}$. For the remainder of this discussion we assume a stellar mass of 30\,M$_{\odot}$, the L1 point separation of which is 17.8\,R$_{\odot}$, and a stellar radius of 17\,R$_{\odot}$ such that the supergiant almost fills its Roche Lobe. Under these parameters the orbit of IGR J16418$-$4532 is characterised as circular, at an approximately edge on inclination with a radius of 31.9\,R$_{\odot}$ and a NS orbital velocity of 432\,km\,s$^{-1}$. 

While the choice of these parameters may seen somewhat overly arbitrary we wish to point out that the defining parameter in our remaining considerations, due to its strong influence on the size of the accretion radius, is the velocity of the stellar wind at the orbital separation of the neutron star. Following previous works on the properties of the stellar wind in IGR J16418$-$4532 (i.e. \citet{2012MNRAS.419.2695R}, \citet{2010MNRAS.408.1540D}) we adopt the velocity law of \citet{1975ApJ...195..157C} to describe the stellar wind velocity for both the smooth and clumped wind components, namely

\begin{align}
	v(r) = v_{\infty} ( 1 - 0.9983 \frac{R_{Sg}}{r} )^\beta
	\label{eq:vlaw}
\end{align}	

{\noindent The stellar mass influences this relationship through the derived value of the orbital separation $r$, and the stellar radius influences through the value of $R_{OB}$. However equally influential are the values of the terminal wind velocity $v_{\infty}$ and the powerlaw index $\beta$. Through modelling of the flare luminosity distribution as observed by \emph{Swift}/XRT, \citet{2012MNRAS.419.2695R} place $v_{\infty}$ in the range 800 to 1300\,km\,s$^{-1}$ and $\beta$ between 0.8 and 1.3. The value of the stellar wind velocity at the orbital separation calculated with the above stellar parameters and $v_{\infty}$\,$=$\,1000\,km\,s$^{-1}$, $\beta$\,$=$\,1 is 470\,km\,s$^{-1}$. However by varying the $v_{\infty}$ and $\beta$ parameters within their limits the stellar wind velocity occupies the range $\sim$370 to 610\,km\,s$^{-1}$. Similarly if we select a different set of allowed stellar parameters (M$_{Sg}$\,$=$\,40M$_{\odot}$, R$_{Sg}$\,$=$\,20R$_{\odot}$) and calculate the stellar wind velocity for the same range of allowed $v_{\infty}$ and $\beta$ values the wind velocity at the orbital separation is in the range $\sim$340 to 560\,km\,s$^{-1}$. Hence the largest source of uncertainty in the stellar wind velocity comes from the poorly constrained wind parameters and not the unknown stellar parameters. For the remainder of this discussion we therefore adopt the stellar parameters outlined above and calculate the relative velocity of the NS orbital and stellar wind radial velocities as $v_{rel} = (v_{NS}^{2} + v_{w}^{2})^{\frac{1}{2}}$. Uncertainties in derived physical values are then discussed in terms of the range of possible stellar wind parameters.

\subsection{Eclipse Emission}

Assuming that the cut off in flux observed by \emph{INTEGRAL}/IBIS was due to the eclipse ingress this provides an accurate ephemeris of MJD\,56171.537$\pm$0.0058 for the start of the X-ray eclipse. Combining this with the first \emph{XMM-Newton} observations (MEO) allows a firm lower limit, of 0.583\,days ($\Delta\phi$ $=$ 0.156), to be placed on the duration of the X-ray eclipse in IGR J16418$-$4532.

The detection of X-ray flux from HMXB systems whilst the X-ray source is eclipsed by the supergiant companion is a common phenomenon amongst such sources, being detected in both classical Sg-XRBs (e.g. Cen X-3, \citealt{2011ApJ...737...79N}) and SFXTs (e.g. IGR J16479$-$4514, \citealt{2008MNRAS.391L.108B}). The detected 0.5$-$10\,keV flux of 2.8$\times$10$^{-13}$\,erg cm$^{-2}$ s$^{-1}$ observed during the MEO corresponds to a luminosity of 6.3$\times$10$^{33}$\,erg s$^{-1}$ at the assumed source distance, ruling out direct emission from the companion star as the source of the detected emission. Instead the detections during eclipse are usually interpreted as resulting from the reprocessing of the intrinsic flux from the NS through the dense wind of the supergiant. In the MEO emission lines at energies consistent with atomic transitions from both neutral (6.4\,keV) and ionised (6.67\,keV) Fe were detected. The lack of Fe line features in the PEO indicate that the iron emission does not originate from close to the NS, but in extended regions throughout the supergiant atmosphere, with the relative equivalent widths of the lines (0.63 and 0.33\,keV for the 6.4 and 6.67\,keV lines respectively) suggesting that neutral Fe is the more abundant state.  

\subsection{Out of Eclipse Emission}

%%%%%%%%%%%%%%%%%%%%%%%%%%%%%%%%%%%%%%%%%%%%%%%%%%%%%%%%%%%%%%%%%%%%%%%%%%%%%%%%%%%%%%%%%%%%%%%%%%%%%%%%%%%
%%% LSidoli 17 Jan 2013 - estimates of the wind density in igr16418 - DISCUSSION SECTION 
%%%%%%%%%%%%%%%%%%%%%%%%%%%%%%%%%%%%%%%%%%%%%%%%%%%%%%%%%%%%%%%%%%%%%%%%%%%%%%%%%%%%%%%%%%%%%%%%%%%%%%%%%%%
%
Following the method of \citet{2013MNRAS.tmp..482S}, it is possible to use both \emph{XMM-Newton} observations (MEO and PEO) to probe the supergiant total mass loss rate and the stellar wind density at the orbital separation. Using the system parameters discussed above gives an orbital separation of 2.2$\times$10$^{12}$\,cm. Additionally the average flux detected whilst the NS was eclipsed in the MEO (2.8$\times$10$^{-13}$\,erg\,cm$^{-2}$\,s$^{-1}$) is $\sim$0.56\% of the average PEO flux detected outside of the X-ray dip and flare regions ($\sim$\,5.0$\times$10$^{-11}$\,erg\,cm$^{-2}$\,s$^{-1}$). This allows the wind density to be estimated, assuming that the X--ray emission detected during the eclipse is produced by X--ray scattering, as $n_{\rm w} (a)$ = $0.0056 / (a \sigma_{\rm T})$, where  $\sigma_{\rm T}$ is the Thomson scattering cross section and {\em a} is the orbital separation \citep{1992ApJ...389..665L}.
%
% ---------------------------------------------------------------------------------------------------------------
% Comment: 0.0056 derives from the ratio: 2.8E-13 (MEO) / 5E-11 (average PEO flux, not in dip, not in flares...)
% ---------------------------------------------------------------------------------------------------------------
%

The wind density at the orbital separation, $\rho$$_{\rm w}(a)$, is calculated as  6.3$\times$10$^{-15}$\,g\,cm$^{-3}$. Assuming that the wind has a spherical geometry, and  $\beta$=1 \citep{1975ApJ...195..157C}, the ratio between the wind mass-loss rate and the terminal velocity can be calculated as $\dot{M}_{w}/ v_{\infty}$=4\,$\pi$\,$a\,(a-R_{opt}$)\,$\rho_{\rm w}(a)$, giving $\dot{M}_{w}/ v_{\infty}$=9.3$\times$10$^{-18}$\,M$_{\odot}$\,km$^{-1}$. If the wind terminal velocity is in the range 800$-$1300\,km\,s$^{-1}$ \citep{2012MNRAS.419.2695R}, the mass loss rate via the stellar wind lies in the range $\dot{M}_{w}$\,=\,(2.3$-$3.8)$\times$10$^{-7}$\,M$_{\odot}$\,yr$^{-1}$. Assuming direct accretion from the wind, the accretion rate onto the NS can be estimated as $\dot{M}_{acc}$\,$=$\,(($\pi R^{2}_{acc} / 4 \pi a^{2}$)$\dot{M}_{w}$), where R$_{acc}$ is the accretion radius and is given by 2GM$_{NS}$/${v_{rel}^{2}}$. This implies an accretion rate at the orbital separation of between $\dot{M}_{\rm acc}$\,$\sim$\,(9.7 - 5.4)$\times$10$^{15}$\,g\,s$^{-1}$, corresponding to X-ray luminosities in the range, L$_{\rm X} =$\,(1.8$-$1.0)$\times$10$^{36}$\,erg\,s$^{-1}$.

The average, unabsorbed flux detected during the PEO, outside of the X-ray dip and flare, corresponds to a luminosity of 2.0$\times$10$^{36}$\,erg\,s$^{-1}$ at a distance of 13\,kpc, implying that the estimated wind density at the orbital separation, for the range of parameters assumed here, is able to generate the accretion luminosity observed outside of the eclipse in IGR J16418$-$4532. This is in contrast to the case of IGR J16479$-$4514 where the calculated stellar wind densities produced an X-ray luminosity two orders of magnitude greater than that observed, suggesting the need for a damping mechanism to reduce the effective mass accretion rate in this system \citep{2013MNRAS.tmp..482S}. Such a mechanism would appear to not be required at nominal luminosities in the case of IGR J16418$-$4532.

\label{sect:OEE}

\subsubsection{n$_{H}$ variations}

The evolution of the spectral parameters and source flux observed during the PEO (Fig. \ref{fig:OBS2specparamfixedGamma}) is intriguing, with evidence of a very variable accretion environment being experienced by the NS in IGR J16418$-$4532. Two striking features are observed during the course of the observation, the first being the deep dip observed in the detected flux that occurs between $\sim$\,t\,$=$\,800 and 2500\,s which is discussed separately below. The second is the pronounced increase and evolution in absorption observed between t\,$\sim$\,7000\,s and t\,$\sim$\,10,000\,s.  The absorption first shows a sharp rise to $\sim$\,10$^{23}$\,cm$^{-2}$ from the base level of $\sim$\,7\,$\times$\,10$^{22}$\,s followed by a second rapid step up to 1.7\,$\times$\,10$^{23}$\,cm$^{-2}$ approximately 1500\,s later. After an additional 1500\,s the absorption then rapidly decreases back to the base level observed in the earlier parts of the observation whilst, almost simultaneous, the flux begins to rise into the flaring activity observed at the end of the observation.

To understand this evolution we first consider the variation in n$_{H}$ observed as the NS moves through its orbit.  \citet{2012MNRAS.420..554S} proposed that the NS in IGR 16418$-$4532 could be accreting through Transitional Roche Lobe Overflow (TRLO) due to the systems narrow orbit and the flux variations observed in \emph{XMM-Newton} data taken in 2011. Such a transitional scenario was investigated by \citet{1991ApJ...371..684B} where model orbital n$_{H}$ profiles were produced for wind accretion powered binaries with narrow, circular orbits (Figure 2 of that work). The overall shape of the profile is dominated by the line of sight through the bow shock about the NS with an additional, prominent component being generated by the tidal gas stream extending from the L1 point. The PEO was separated from the middle of the eclipse by an orbital phase interval of $\sim$\,0.2 which should, according to \citet{1991ApJ...371..684B}, result in a smooth decline in n$_{H}$. This suggests that the increase in n$_{H}$ observed is not generated by a coherent structure in the NS environment (i.e. a tidal stream) but is instead more likely to be generated by an individual dense clump of material travelling through the line of sight to the NS at this time. Significant variation in the photon index of the powerlaw was not observed during the period of enhanced n$_{H}$ and the unabsorbed flux remained at a constant level during this time (Fig. \ref{fig:OBS2nHunabsf}, red line). This is likely a signature of absorption of a constant intrinsic flux by optically thin material, as opposed to an optically thick/ionised absorber which might be expected in the case of obscuration by a coherent structure such as a tidal stream. We note that sensitive, soft X-ray observations at a later orbital phase would be required to further test for the presence of a tidal stream arising from TRLO in IGR J16418$-$4532.

A dynamical estimate of the extent of the over dense region required to generate the excess n$_{H}$ observed can be made by utilising the calculated relative velocity between the NS orbital motion and the radial stellar wind (see above). The increase in n$_{H}$ is observed for $\sim$\,2500\,s which relates to the passage of a wind clump of radius $\sim$\,8\,$\times$\,10$^{10}$\,cm travelling at a velocity of 470\,km\,s$^{-1}$. This radius is consistent with the range of clump radii derived from the theoretical considerations of \citet{2009MNRAS.398.2152D} and also similar to the observational results of \citet{2011A&A...531A.130B} who explained an outburst of IGR J18410$-$0535 as resulting from the accretion of a clump of approximate radius 8\,$\times$10\,$^{11}$\,cm. Therefore we conclude that this absorption feature likely results from an optically thin clump of stellar wind material obscuring the intrinsic emission from the neutron star.

\subsubsection{Absorption and flaring link?}

The temporal relationship between the n$_{H}$ increase and the onset of the flaring activity at the end of the PEO is of interest. Figure \ref{fig:OBS2nHunabsf} shows the n$_{H}$ evolution together with the unabsorbed 0.5\,$-$\,10\,keV flux derived from the fixed $\Gamma$ fit to each spectrum. It can be seen that the unabsorbed flux only begins to rise above the base level in the observation as the n$_{H}$ returns to its pre-enhancement level. The timing of these two events is intriguing, but on the basis of this dataset alone no conclusive physical connection can be drawn between these features and instead they may simply be generated by two separate, unrelated events that randomly occurred sequentially. Given the tighter constraints on IGR J16418$-$4514s orbital geometry through the X-ray eclipse however, a consideration as to whether such a delay could be generated through physical means and indicate a causal link between the n$_{H}$ and flux evolutions observed can be undertaken. 

The lack of an overlap in the evolution of these parameters appears to be in contradiction to the best example of clump accretion in an SFXT as presented in \citet{2011A&A...531A.130B}. In this work an outburst of IGR J18410$-$0535 was described through the accretion of a dense clump of wind material. The n$_{H}$ was observed to rise throughout the entire flux evolution, only returning to its pre-flare level when the flux also returned to pre-flare level, suggesting that there was excess material in the immediate vicinity of, and line of sight to, the NS throughout the duration of the flare. However in IGR J16418$-$4532 it appears that if the denser clump material is responsible for generating the subsequent flare then the clump was only in our line of sight to the neutron star prior to the flux increase. This discrepancy seems somewhat unintuitive but can start to be understood from the more detailed knowledge of the orbital geometry of the IGR J16418$-$4532 system as a result of the X-ray eclipse. Using Fig. \ref{fig:pfold} we can centre the eclipse at an orbital phase of approximately $\phi$\,$=$\,0.55 meaning that in the orbital phase range $\phi$\,$=$\,0.3 to 0.8 the neutron star is more distant than the supergiant companion. Hence in this phase range, which the PEO inhabits, stellar wind clumps that are accreted will only be in our line of sight to the neutron star as they are approaching and interacting with it. As the clump moves beyond the X-ray emitting region the absorption detected along our line of sight would then drop rapidly, as is observed in Fig. \ref{fig:OBS2nHunabsf}, as the clump moves away behind the neutron star and we are therefore unable to continue to detect its presence. 

To further investigate whether this geometrical effect could be behind the observed evolution we consider the free fall time of stellar wind material from the accretion radius. We again assume the accretion radius is given by R$_{a}$\,$=$\,2GM$_{NS}$/${v_{rel}^{2}}$, and the free fall time is given by t$_{ff}$\,$=$\,(R$_{a}^{3}$/2GM$_{NS}$)$^{\frac{1}{2}}$. Under these assumptions and using the previously derived system parameters, the accretion radius is calculated as $\sim$\,9$\times10^{10}$\,cm and the associated free fall time is $\sim$1500\,s. This value should be considered a lower limit on the true infall time however as the descending material does not fall through the gravitational potential unimpeded, likely becoming a turbulent flow at some point which will increase the infall time. Given that the calculated free fall time is comparable to the observed delay between the onsets (and peaks) of the n$_{H}$ and unabsorbed flux evolutions, Fig. \ref{fig:OBS2nHunabsf}, and the X-ray generation region is close to the neutron star, it is possible that in this case the dense wind clump passed through the accretion sphere before the additional accreted material had reached the X-ray generation region, hence creating the delay observed in Fig. \ref{fig:OBS2nHunabsf}.  

Through considerations of the orbital geometry of IGR J16418$-$4532, the temporal coincidence (of the order of the free fall time from $R_{a}$) and approximately similar durations of the n$_{H}$ and flux evolutions, we suggest that a clump of over dense, optically thin wind material approaching the neutron star at a distance comparable to the accretion radius would be consistent with the production of both features. In this case IGR J16418$-$4532 is showing signatures of accretion through the `clumpy wind' model \citep{2005A&A...441L...1I,2007A&A...476..335W} viewed at an approximately edge-on inclination. However we again stress that at the current time it is not possible to definitively conclude that these two features are causally related and that they might instead be caused by two physically unrelated events. Long term monitoring of IGR J16418$-$4532 across all orbital phases would allow firmer conclusions to be drawn as to any physical link between these features through the detection, or otherwise, of additional n$_{H}$ - flux delays and their distribution in orbital phase. If it were the case that these features are physically related, it cannot currently be discerned if the difference in the spectral evolution observed in IGR J16418$-$4532 compared to the example of IGR J18410$-$0535 \citep{2011A&A...531A.130B} is due to the same geometrical effects, as the orbital configuration of IGR J18410$-$0535 is currently unknown. 

Finally there is some evidence of a slow decrease in the base n$_{H}$ level at times greater than t\,$=$\,10,000\,s which could be the result of increased X-ray photoionisation of the absorbing material by the enhanced X-ray flux produced during the flare \citep{1977ApJ...211..552H}, however the observation ended before this effect could be more firmly identified. 

\subsubsection{The nature of the `dip'}

We now consider the properties of, and possible physical mechanisms behind, the X-ray dip observed towards the start of the PEO. Due to the unfortunate lack of EPIC-pn coverage of the majority of the dip region we were unable to perform a detailed spectral analysis of the ingress and mid-dip regions, as shown by the large error bars and poor $\hat{\chi}^{2}$ values in Figs. \ref{fig:OBS2specparam} and \ref{fig:OBS2specparamfixedGamma}, but did achieve more sensitive coverage during the dip egress. The main observable features of the X-ray dip are:

\begin{description}
	\item $-$ a total duration of $\sim$\,1500\,s between t\,$=$800 and 2300\,s that is comparable to the spin period of the neutron star
	\item $-$ a sharp drop in flux lasting less than 300\,s at the start of the feature
	\item $-$ a steady, low level flux lasting for $\sim$\,900\,s with an unabsorbed flux of 2.0$\times$10$^{-11}$\,erg\,cm$^{-2}$\,s$^{-1}$ (0.5 $-$ 10\,keV) corresponding to a luminosity of 4.1$\times$10$^{35}$\,erg\,s$^{-1}$ at a distance of 13\,kpc
	\item $-$ a slightly less sharp return to the pre-dip flux level over $\sim$300\,s with some evidence of the n$_{H}$ decreasing back to the pre-dip level as the flux recovers (see Fig. \ref{fig:OBS2nHunabsf})
	\item $-$ a decrease in flux observed across the full 0.2$-$10\,keV energy range during which spectral changes could not be identified
	\item $-$ a smooth spectrum with no evidence of Fe emission lines.
\end{description}

The presence of X-ray intensity dips, also referred to as `off-states', in the light curves of wind-fed HMXB pulsars is a rare phenomenon, having only been observed in the classical Sg-XRBs Vela X-1 \citep{2008A&A...492..511K}, Centaurus X-3 \citep{2011ApJ...737...79N}, 4U 1907$+$09 \citep{1998ApJ...496..386I} and GX 301$-$2 \citep{2011A&A...525L...6G}. \citet{2012A&A...544A.118B} also reported a similar feature in the non-pulsating, candidate SFXT IGR J16328$-$4726 but a detailed analysis and characterisation was prevented due to statistical constraints. Different mechanisms have been proposed to explain the X-ray dips observed in the above sources. Here we consider their applicability to IGR J1618$-$4532, whilst also noting that additional, as yet unexplained, mechanisms may be acting to create this feature.

The eclipsing, Roche-Lobe filling Sg-XRB system Cen X-3 displayed several X-ray dips during the course of a long \emph{Suzaku} observation that covered almost an entire binary orbit \citep{2011ApJ...737...79N}. The dips were observed up to energies of 40\,keV and the spectra characterised by a typical accreting HMXB pulsar continuum with multiple Fe emission lines that were detected throughout the observation. The equivalent widths (EW) of the Fe lines showed significant enhancement during both the dip and binary eclipse regions of the exposure. By following the evolution of the Fe line EWs and the amount of absorbing material in the line of sight (n$_{H}$ values increased by a factor of 100 during these times), the authors concluded that the dips were most likely caused by the obscuration of the neutron star by dense structures in the outer edge of the accretion disc in the system. In this way Cen X-3 is a HMXB analogy of the `dipper' class of LMXBs. Obscuration due to dense structures in the vicinity of the neutron star is unlikely in the case of IGR J16418$-$4532 however. Observationally the detection of prominent Fe emission during the MEO but not during the dip suggests that obscuration by dense, compton thick material is not the cause of the intensity dip observed in these observations. Additionally it is not clear how such dense material could inhabit the IGR J16418$-$4532 system as the supergiant is not believed to overfill its Roche-Lobe and therefore an accretion disc would not be present. In the TRLO regime such dense material could be found as a result of either the gravitationally focused `tidal stream' coming from the L1 point or as a result of a transient accretion disc that may form. However, as discussed previously, the orbital phase location of the dip is inconsistent with what would be expected if it were to originate from a tidal stream \citep{1991ApJ...371..684B}. \citet{2012MNRAS.420..554S} suggested the presence of a transient accretion disc to explain the quasi-periodic flares observed in a previous observation of IGR J16418$-$4532, however as no such features are present in these observations it is unlikely that a transient accretion disc was present to obscure the NS at this time. We note however that these observations cover an insufficient fraction of orbital phase to consider fully whether TRLO is occurring in IGR J16418$-$4532.    

The signature of a typical stellar wind clump obscuring the neutron star in IGR J16418$-$4532 is believed to have been observed later in the PEO, as discussed previuosly, and displayed a very different evolution to that observed during the dip feature. Hence it is also unlikely that the passage of a single stellar wind clump through the line of sight is responsible for the observed dip. Finally the passage of ionised material through the line of sight to the neutron star could cause a strong absorbing effect across a wide energy range as observed. Again however there was insufficient signal to test for this effect spectrally and the lack of Fe emission lines in the X-ray spectrum of the dip suggests this is an unlikely cause of the observed feature.

The alternative explanation given for the presence of X-ray intensity dips or `off-states' in the light curves of wind-fed SgXRBs is the action of a barrier to the accretion flow. \citet{2008A&A...492..511K} invoked centrifugal barriers to explain the off-states observed in Vela X-1, where the source fell below the detection threshold in \emph{INTEGRAL}/IBIS data, via the onset of the supersonic propellor mechanism \citep{1975A&A....39..185I} to inhibit accretion during these times. More recently however \citet{2011A&A...529A..52D} used sensitive, broadband \emph{Suzaku} observations to identify the continued presence of X-ray pulsations during the dips of Vela X$-$1 and argued that the dips are the result of a magnetic barrier that reduces the emitted X-ray flux as direct accretion is impeded by the neutron star magnetosphere (subsonic propellor regime, see \citealt{2008MNRAS.391L.108B}). In this state accretion continues at a lower level through the Kelvin-Helmholtz Instability (KHI) which generates the lower flux levels detected during the X-ray dips. \citet{2011A&A...529A..52D} used the detailed accretion scenario model of \citet{2008MNRAS.391L.108B} to infer that a neutron star with a magnetic field strength in the range of a few 10$^{13}$ to $\sim$1\,$\times$\,10$^{14}$\,G is required to produce the X-ray luminosity observed under this model and argue that X-ray intensity dips in wind-fed HMXBs are evidence of a highly magnetised neutron star being present in the system. The same authors also found similar results for the HMXB 4U 1907$+$09 \citep{2012A&A...548A..19D}.

\citet{2013MNRAS.428..670S} however, proposed that the X-ray dips observed in Vela X-1, GX 301$-$2 and 4U 1907$+$09 were instead caused by a switch in the polar beam configuration due to a variation in the optical depth of the accretion column. In this model both states occur within a `quasi-spherical' subsonic accretion regime \citep{2012MNRAS.420..216S}. The Compton cooling regime occurs when the source luminosity is in excess of $\sim$5$\times$10$^{35}$\,erg\,s$^{-1}$ and the optical depth of the accretion column is high, creating a fan beam due to high levels of lateral scattering. The fan beam effectively cools plasma in the equatorial regions of the magnetosphere through Compton processes, allowing material to enter the magnetosphere through Rayleigh-Taylor instabilities. It is this state that generates the `normal' luminosity level of the lower luminosity, persistent wind-fed HMXBs (i.e. Vela X-1). The radiative cooling regime occurs when the luminosity is less than $\sim$5$\times$10$^{35}$\,erg\,s$^{-1}$, causing a drop in the optical depth of the accretion column and the onset of pencil beam emission. In this configuration the equatorial region of the magnetosphere is not illuminated and plasma can only thread onto field lines at a lower rate through radiative plasma cooling, which is independent of the level of illumination from the central source. The authors argue it is the transition into this radiatively cooled accretion state from the Compton cooled state that generates the observed X-ray dips. Sources would only move into the direct accretion scenario (supersonic Bondi accretion) when the source luminosity increased to greater than several 10$^{36}$\,erg\,s$^{-1}$ and in-falling material is rapidly cooled by Compton processes before reaching the magnetosphere (for a full description of the quasi-spherical accretion model see \citet{2012MNRAS.420..216S} and \citet{2013MNRAS.428..670S}).          

Given the  similarities in the observable features and luminosities of the X-ray dips observed in Vela X-1 and 4U 1907$+$09 to the dip observed here, we consider the applicability of the \citet{2011A&A...529A..52D} and \citet{2013MNRAS.428..670S} interpretations of the origins of X-ray dips to IGR J16418$-$4532. We first follow the considerations of \citet{2011A&A...529A..52D}, using the theoretical framework of \citet{2008MNRAS.391L.108B}, to calculate the magnetic field strength required to produce the observed dip luminosity (4.1$\times$10$^{35}$\,erg\,s$^{-1}$) for the range of stellar wind parameters considered in Section \ref{sect:OEE}. We use Equation 21 of \citet{2008MNRAS.391L.108B} and the conventions therein, namely:

\begin{equation}
	\begin{aligned}
		L_{KH} = &1.8 \times 10^{35} \eta_{KH} P^{-1}_{s3} R^{3}_{M10} \dot{M}_{-6} a^{-2}_{10d} v^{-1}_{8} \\
		&[1 + 16R_{a10}/(5R_{M10})]^{3/2} \frac{\sqrt{\rho_{i}/\rho_{e}}}{(1 + \rho_{i}/\rho_{e})}  \textnormal{erg\,s$^{-1}$} 
	\end{aligned}
\end{equation}

{\noindent Here the $\rho_{i}/\rho_{e}$ term is the ratio of the plasma densities inside and outside of the magnetosphere and $R_{M10}$ is the magnetic radius in units of 10$^{10}$\,cm and defined as:}

\begin{equation}
	\begin{aligned}
		R_{M} = 2 \times 10^{10} a_{10d}^{4/7} \dot{M}_{-6}^{-2/7} v_{8}^{8/7} \mu_{33}^{4/7} \textnormal{cm} \\
	\end{aligned}
\end{equation}

{\noindent where $\mu_{33}$ is the magnetic moment in units of 10$^{33}$\,G\,cm$^{3}$. We calculate the magnetic field required to produce the dip luminosity in the cases of $\rho_{i}/\rho_{e} =$ 1.0, 0.5 and 0.1 where a $\rho_{i}/\rho_{e}$ value of 1 gives an upper limit on the mass flow generated through the Kelvin-Helmholtz instability.} 

Using the calculated magnetic field strengths we then estimate the critical accretion rate defining the transition between the subsonic propellor and direct accretion regimes. According to \citet{2008MNRAS.391L.108B} this critical accretion rate is given by:

\begin{equation}
	\begin{aligned}
		\dot{M}_{lim-6} =& 2.8 \times 10^{2} P^{-3}_{s3} a^{2}_{10d} v_{8} \\ 
		 &R^{5/2}_{M10} [1 + 16R_{a10}/(5R_{M10})]^{-3/2} \textnormal{g\,s$^{-1}$}
	\end{aligned}
\end{equation}

{\noindent These critical values are compared to the accretion rates of 2.2\,$\times$10$^{15}$\,g\,s$^{-1}$ and 1.1\,$\times$10$^{16}$\,g\,s$^{-1}$ (using $L_{X} = GM_{NS} \dot{M}_{capt} / R_{NS}$) required to generate the observed dip luminosity of 4.1$\times$10$^{35}$\,erg\,s$^{-1}$ and the pre/post-dip luminosity of $\sim$2$\times$10$^{36}$\,erg\,s$^{-1}$ respectively to evaluate if a transition between the accretion regimes of this model could have occurred. Table \ref{tab:Bfields} outlines the results of these calculations for the range of possible stellar wind velocities considered in Section \ref{sect:OEE}}.

\begin{table*}
\caption{Calculated magnetic fields required to generate the observed dip luminosity in the `Subsonic Propellor' accretion mode according to \citet{2008MNRAS.391L.108B}. The wind condition parameter space is drawn from the limits on the terminal wind velocity range of \citet{2012MNRAS.419.2695R} and the total stellar wind mass loss rate calculated using the $\dot{M}_{w}/ v_{\infty}$=9.3$\times$10$^{-18}$\,M$_{\odot}$\,km$^{-1}$ parameter for each terminal wind velocity (see Sect. \ref{sect:OEE}). The `Transition?' columns indicate whether a transition from the subsonic propellor regime to the direct accretion regime is likely to have occurred for the X-ray dip and flare observed in the observation. The accretion rates required to produce the dip, pre/post-dip and peak flare luminosities are 2.2\,$\times$10$^{15}$\,g\,s$^{-1}$, 1.1\,$\times$10$^{16}$\,g\,s$^{-1}$ and 2.8\,$\times$10$^{16}$\,g\,s$^{-1}$ respectively.} 
\begin{center}
\begin{tabular}{|c|c|c|c|c|}
\hline
\multicolumn{1}{|c|}{$\rho_{i}/\rho_{e}$} & \multicolumn{1}{|c|}{B(G)} & \multicolumn{1}{|c|}{$\dot{M}_{lim}$ (g\,s$^{-1}$)} & \multicolumn{1}{|c|}{Dip Transition?} & \multicolumn{1}{|c|}{Flare Transition?} \\ \hline 
\multicolumn{5}{|l|}{$v_{\infty}$\,$=$\,800\,km\,s$^{-1}$, $\dot{M}_{w}$\,$=$2.3$\times$10$^{-7}$ M$_{\odot}$\,yr$^{-1}$, $v_{rel}$\,$=$\,571\,km\,s$^{-1}$} \\
 1.0 & 2.8$\times$10$^{13}$ & 1.2$\times$10$^{17}$ & N &N \\ 
 0.5 & 3.0$\times$10$^{13}$ & 1.4$\times$10$^{17}$ & N & N \\ 
 0.1 & 5.3$\times$10$^{13}$ & 5.2$\times$10$^{17}$ & N & N \\ \hline
\multicolumn{5}{|l|}{$v_{\infty}$\,$=$\,1000\,km\,s$^{-1}$, $\dot{M}_{w}$\,$=$2.9$\times$10$^{-7}$ M$_{\odot}$\,yr$^{-1}$, $v_{rel}$\,$=$\,637\,km\,s$^{-1}$} \\
 1.0 & 3.2$\times$10$^{13}$ & 3.2$\times$10$^{17}$ & N & N \\ 
 0.5 & 3.4$\times$10$^{13}$ & 3.8$\times$10$^{17}$ & N & N \\ 
 0.1 & 6.1$\times$10$^{13}$ & 1.4$\times$10$^{18}$ & N & N \\ \hline
\multicolumn{5}{|l|}{$v_{\infty}$\,$=$\,1300\,km\,s$^{-1}$, $\dot{M}_{w}$\,$=$3.8$\times$10$^{-7}$ M$_{\odot}$\,yr$^{-1}$, $v_{rel}$\,$=$\,746\,km\,s$^{-1}$} \\
 1.0 & 4.1$\times$10$^{13}$ & 1.5$\times$10$^{18}$ & N & N \\ 
 0.5 & 4.3$\times$10$^{13}$ & 1.8$\times$10$^{18}$ & N & N \\ 
 0.1 & 7.6$\times$10$^{13}$ & 6.4$\times$10$^{18}$ & N & N \\ \hline
 \end{tabular}
 \label{tab:Bfields}
 \end{center}
 \end{table*} 
 
It is seen from Table \ref{tab:Bfields} that transitions between these two accretion regimes cannot be responsible for the observed dip in the light curve of IGR J16418$-$4532 as the critical transition accretion rates are in excess of the dip, pre/post-dip and peak-flare (2.8\,$\times$10$^{16}$\,g\,s$^{-1}$) accretion rates required to generate the observed luminosity levels across the whole parameter space investigated. Hence this suggests that a transition between the `subsonic propellor' and direct accretion regimes is not causing the X-ray dip observed in IGR J16418$-$4532.

Given the earlier arguments against obscuration as the source of the dip and the incompatibility of the required accretion rates for a subsonic propellor to direct accretion transition, we investigate a state change within the `quasi-spherical' accretion model \citep{2013MNRAS.428..670S} as a possible cause of the X-ray dip in IGR J16418$-$4532. As IGR J16418$-$4532 appears to stay in a subsonic quasi-spherical accretion regime throughout the observation this suggests the X-ray dip may be generated through a transition between the higher luminosity fan beam and lower luminosity pencil beam dominated states of the model. Under this scenario we can estimate the magnetic field strength of the NS through the luminosity detected during the dip using Equation 22 of \citet{2013MNRAS.428..670S}, namely:

\begin{equation}
	\begin{aligned}
		L_{X,rad} \approx 10^{35} \mu^{7/33}_{30} \textnormal{erg\,s$^{-1}$}
	\end{aligned}
\end{equation} 

where $\mu_{30}$ is the magnetic moment in units of 10$^{30}$ G\,cm$^{3}$. The dip luminosity of 4.1$\times$10$^{35}$\,erg\,s$^{-1}$ corresponds to a NS B-field of $\sim$2$\times$10$^{14}$\,G, indicating that the presence of a highly magnetised NS in a SFXT would be required by this model. Additionally the shell of material that would be surrounding the NS under this model could be providing a significant contribution to the high base level of absorption observed in the PEO ($\sim$7$\times$10$^{22}$\,cm$^{-2}$). As the derived B-field value is determined by the luminosity of the source during the X-ray dip we consider the effect of the unknown source distance on this interpretation. The confirmation of the companion as a supergiant requires the source to be at a large distance for all possible stellar radii \citep{2008A&A...484..783C}, however a distance of 13\,kpc is still somewhat arbitrary and is used here to maintain consistency with previous works in the literature. If IGR J16418$-$4532 is located at a distance of 10\,kpc (for example) then the dip luminosity would be 2.4$\times$10$^{35}$\,erg\,s$^{-1}$ and the implied NS B-field would have a value of 1.6$\times$10$^{13}$\,G. Whilst in this case the B-field doesn't reach magnetar field strengths it is nevertheless in the highly magnetised regime, being in excess of the nominal $\sim$10$^{12}$\,G value observed in many HMXB pulsars. 

Due to the lack of EPIC-pn coverage during the early regions of the dip it is difficult to evaluate what may have caused the on-set of an accretion regime transition for the duration of the dip. However the hydrodynamical simulations of \citet{2012MNRAS.421.2820O} showed that, whilst on average the velocity of the stellar wind follows the velocity law stated above (Eq. \ref{eq:vlaw}), strong velocity jumps over small radial extents are also present. These jumps, which can have negative gradients, result from unstable growth in the line-driven wind and are at their most prominent at separations within a few stellar radii of the supergiant surface. As a result of its short orbit the NS in IGR J16418$-$4532 is occupying the most turbulent region of the supergiant wind and an accretion regime transition may have been triggered by a strong, localised variation in the stellar wind environment interacting with the magnetosphere.

The previous paragraphs have outlined the applicability of the mechanisms currently proposed to generate the X-ray dips observed from some SgXRBs. Through these considerations we conclude that the most applicable of these mechanisms is that of an accretion regime transition between the two subsonic states of the `Quasi-spherical accretion' model \citep{2012MNRAS.420..216S}. An important implication of this model is that of the requirement of a highly magnetised neutron star to be present within the IGR J16418$-$4532 system. Caution must be taken with these interpretations however due to the large number of assumptions required to perform these calculations. The detection and characterisation of additional dips is required, along with independent constraints on the orbital, stellar and stellar wind parameters of IGR J16418$-$4532, to allow the testing of the applicability of all of the possible causes of the dip to the IGR J16418$-$4532 system under more stringent physical constraints. At the current time it is therefore not possible to definitively identify the physical mechanisms that create the dip feature.

\subsection{The nature of IGR J16418$-$4532}

The strongly variable stellar wind environment being experienced by the NS in IGR J16418$-$4532 may also explain the origin of the enhanced dynamic range ($> 10^{2}$) observed in this system compared to classical Sg-XRBs. Many Sg-XRBs have longer orbits and should therefore occupy less turbulent regions of their supergiants atmosphere resulting in a lower level of variability. The sharp velocity and density variations near to the surface of the supergiant in IGR J16418$-$4532 however may be helping to generate the observed SFXT level of variability in the system through their influence on the size of the accretion radius long with the achieved mass capture rate. 

The transient nature of IGR J16418$-$4532 is defined more by the large source distance of $\sim$13\,kpc however. Figure \ref{fig:pfold} shows that outside of the eclipse IGR J16418$-$4532 is detected as a weak persistent source in the \emph{INTEGRAL}/IBIS 18$-$60\,keV band (where most SFXTs were discovered) at a count rate of 1.2 counts per second ($\sim$7\,mCrab). If IGR J16418$-$4532 was located at a distance of 2\,kpc however (comparable with Vela X-1, \citealt{1989PASJ...41....1N}) it would be detected as a bright persistent Sg-XRB source with an unusually high dynamic range and an average flux of $\sim$300\,mCrab in the \emph{INTEGRAL} band. Hence it would not be viewed as a transient source. Therefore one may take the view that IGR J16418$-$4532 is a classical Sg-XRB with an enhanced X-ray dynamic range resulting from the turbulent stellar wind environment being experienced by the NS as a result of the short orbit orbital period of the system. Its transient nature is then being generated as a result of its large distance making the persistent emission undetectable in single \emph{INTEGRAL}/IBIS or \emph{Swift}/BAT observations, but where the flares are instead detected as transient outbursts. This is at odds with many other SFXTs where the driver of the extreme level of flux variation is the large scale variation in stellar wind environment experienced by a NS over a longer, more eccentric orbit (e.g. SAX J1818.6$-$1703, \citealt{2009MNRAS.393L..11B}). \citet{2012MNRAS.420..216S} also suggest that the difference between SFXTs and classical wind-fed Sg-XRBs could be understood through the proportion of time spent in the fan beam dominated (higher luminosity) and pencil beam dominated (lower luminosity) regimes within the subsonic, quasi-spherical accretion mode. Classical Sg-XRBs (e.g. Vela X-1) spend the majority of the time in the fan beam dominated regime and occasionally enter the pencil beam regime during X-ray dips. Whereas SFXTs spend the majority of their time in the pencil beam dominated regimes and occasionally enter the fan beam regime during short periods of increased accretion, i.e. outbursts. Given IGR J16418$-$4532s emission history and the observations presented here this further suggests that, fundamentally, IGR J16418$-$4532 is a classical wind-fed Sg-XRB displaying an enhanced dynamic range and is observed as a transient source only due to its large distance.

This difference in the fundamental cause of the variability observed in different SFXT systems is suggestive of the fact that the SFXT class is a phenomenological group defined by the duration and dynamic range of the flares and consists of multiple populations of wind-fed Sg-XRBs whose enhanced level of variability arises from different physical origins. These populations could also be the result of different evolutionary paths followed to reach the current Sg-XRB state as \citet{2011MNRAS.415.3349L} argued that some longer orbital period SFXTs (namely IGR J18418$-$0311 and IGR J11215$-$5952) were most likely to have evolved from OeXRB systems. Such systems, as with the BeXRBs known currently, would likely have had longer, more eccentric orbits which would not change significantly when the main sequence Oe star evolved off the main sequence, thus providing the current orbital configuration inferred in the SFXTs \citep{2011MNRAS.415.3349L}.      

To gain further understanding into the accretion processes occurring and the possible different sub-populations that make up the SFXT class will require high sensitivity X-ray observations covering wide regions of orbital phase. Additionally an accurate characterisation of the supergiant stellar and stellar wind parameters in individual SFXTs will aid in identifying the accretion scenarios occurring under stricter physical constraints. In particular achieving sensitive monitoring of SFXTs over the entire orbital period range is imperative in identifying if there is a systematic change in behaviour between sources whose orbital parameters are more consistent with Sg-XRBs and those whose parameters more closely resemble BeXRBs.

\label{sect:DISC}

\section{Conclusions}

In this work we have presented new combined \emph{INTEGRAL} and \emph{XMM-Newton} observations of the intermediate SFXT IGR J16418$-$4532. We have identified features in the \emph{XMM-Newton} data that are consistent with the presence of dense stellar wind clumps as predicted by the nominal `clumpy-wind' model of SFXT emission along with the first ever detection of an X-ray `dip' in the light curve of a pulsating SFXT. Through considerations of the cause of this dip and the luminosity levels detected we conclude that the most likely cause was the switching of accretion modes from a Compton cooling dominated to a radiative cooling dominated regime within the framework of a subsonic quasi-spherical accretion flow \citep{2012MNRAS.420..216S}. Further observations and characterisation of IGR J16418$-$4532 are required to definitively identify this scenario as the correct explanation of the observed behaviour but, if confirmed, IGR J16418$-$4532 would be required to host a highly magnetised, B $\sim 10^{14} G$, neutron star. In this case the observations presented in this paper would represent the first observational evidence of such an object being hosted in an SFXT. It is likely that the observational properties of IGR J16418$-$4532, that of a persistent wind-fed SgXRB with an enhanced dynamic range, are in part defined by the high level of turbulence in the stellar wind being experienced by the neutron star in its short orbit close to the companion supergiant star. The orbital geometry and flare generation mechanisms observed in IGR J16418$-$4532 represent a departure from the nominal model of SFXT behaviour. This suggests that there may be sub-populations within the SFXT class that, whilst having some similar defining observational characteristics (dynamic range, flare luminosities, stellar companions), generate the observed X-ray behaviour through a variety of different processes.  

\label{sect:CONC}

\section*{Acknowledgements}

The authors wish to thank the anonymous referee for their useful and instructive comments. The authors also wish to thank S. Chaty for his helpful discussion and information on the supergiant spectral type in IGR J16418$-$4532. S. P. Drave acknowledges support from the UK Science and Technology Facilities Council, STFC. M. E. Goossens is supported by a Mayflower scholarship from the University of Southampton. L. Sidoli acknowledges financial support from PRIN INAF 2009. V. A. McBride acknowledges support from the University of Cape Town (UCT) and the National Research Federation (NRF) of South Africa. A. B. Hill acknowledges that this research was supported by a Marie Curie International Outgoing Fellowship within the 7$^{th}$ European Community Framework Programme (FP7/2007--2013) under grant no. 275861. This research has made use of the SIMBAD database, operated at CDS, Strasbourg, France. This research has made use of the IGR Sources page maintained by J. Rodriguez \& A. Bodaghee (http://irfu.cea.fr/Sap/IGR-Sources/).  

%{\bf Please let me know of any financial acknowledgements you would like me to provide here}

\label{lastpage}

\end{document}